\documentclass[preprint,showpacs,12pt]{revtex4}
\usepackage{amsmath,amssymb,amsfonts}
\usepackage{graphicx}
\usepackage{hhline}
\usepackage{bm}
\usepackage{amsmath}
\usepackage{epsfig}
\begin{document}

\title{Low-energy dipole excitations in neon isotopes and $N=16$ isotones within the quasiparticle random phase approximation and the Gogny force}

\author {M. Martini, S. P\'eru and M. Dupuis}

\affiliation{CEA/DAM/DIF, F-91297 Arpajon, France}

\begin{abstract}
Low-energy dipole excitations in neon isotopes and $N=16$ isotones are calculated with a fully consistent axially-symmetric-deformed
quasiparticle random phase approximation (QRPA) approach based on Hartree-Fock-Bogolyubov (HFB) states. 
The same Gogny D1S effective force
has been used both in HFB and QRPA calculations. The microscopical structure of these low-lying resonances,
as well as the behavior of proton and neutron transition densities, 
are investigated in order to determine the isoscalar or isovector nature of the excitations.
It is found that the $N=16$ isotones $^{24}$O, $^{26}$Ne, $^{28}$Mg, and $^{30}$Si are characterized by a similar behavior.
The occupation of the 2s$_{1/2}$ neutron orbit turns out to be crucial, leading to nontrivial transition densities 
and to small but finite collectivity.
Some low-lying dipole excitations of $^{28}$Ne and $^{30}$Ne, characterized by transitions involving the $\nu$1d$_{3/2}$ state, present
a more collective behavior and isoscalar transition densities. A collective proton low-lying excitation 
is identified in the $^{18}$Ne nucleus.

\end{abstract}


\maketitle

\section{Introduction}
Since the beginning of the nuclear physics, dipole excitations of nuclei represented one of the
most important ways to investigate nuclear structure.
This is due to the fact that they can be induced by $\gamma$ radiation.
The first evidence of isovector giant dipole resonance (IVGDR) was obtained in the 1937 \cite{BOT37}.
Extensive experimental and theoretical studies started about ten years later and provided 
simple expressions that systematically reproduce the excitation energy and the strength of the IVGDR in the whole nuclear chart \cite{ww}.
We are currently living in a sort of Renaissance era for this dipole mode since it
plays a crucial role in the study of skin and halo structures in nuclei with large proton-neutron
asymmetry, which can be obtained with radioactive nuclear beams. There is in fact experimental evidence that in nuclei with neutron excess, in addition to the well-known giant dipole resonance,
an accumulation of strength appears at low energies
\cite{Leistenschneider:2001zz,Ryezayeva:2002zz,Hartmann:2004zz,Adrich:2005zz,Schwengner:2008rk,Savran:2008zz,Gibelin:2008zz,Wieland:2009zz,Tonchev:2010zz}.
The nature of this dipole resonance is an open problem.
In light nuclei, several theoretical calculations seem to relate it to nonresonant independent single-particle excitations
of loosely bound neutrons \cite{Paar:2007bk}, but there are some experimental results, for example for the $^{26}$Ne  \cite{Gibelin:2008zz}, which
disagree with this picture. In medium and heavy nuclei this excitation seems to be collective, according to several theoretical calculations \cite{Paar:2007bk},
and it has been interpreted as a resonant oscillation of the neutron skin against the remaining isospin saturated neutron-proton core. It has been called
pygmy dipole resonance (PDR). This denomination is often used to indicate this new collective mode, which in general exhausts a small fraction of the energy-weighted
sum rule (EWSR). Sometimes, on the other hand, it is used in literature to generically indicate low-lying dipole excitations. The degree of collectivity
and its evolution with the mass number is one of the most important open questions, but, as recently explained \cite{Paar:2010ww}, several other open problems
remain in connection with this low-energy dipole excitation. One of them is the isoscalar or isovector nature. Theoretically, it can be investigated by calculating the
neutron and proton transition densities; experimentally, it can be investigated by inducing the same excitation through different processes including electromagnetic processes, 
such as ($\gamma,\gamma'$), ($e,e'$) or Coulomb excitation, on one hand, and strong processes, such as ($\alpha, \alpha'$) or ions collision, on the other hand. Another important problem is the
energy location of this excitation with respect to the neutron threshold, in connection with the question of whether some part of the PDR low-energy tail might be
missing. The evolution of the PDR with neutron-proton asymmetry as well as the existence of a connection between the PDR in stable and exotic nuclei are also of great interest.

Here, we investigate these problems by focusing on neon isotopes and $N=16$ isotones through a fully consistent axially-symmetric-deformed
quasiparticle random phase approximation (QRPA) calculation based on Hartree-Fock-Bogolyubov (HFB) states. The same Gogny D1S effective force \cite{gog1}
 has been used both in HFB and QRPA, which ensures the consistency of the calculations.
This approach is essential in open-shell nuclei, where pairing correlations play an important role.
It also allows the treatment of other isotopic and isotonic chains far from the closed major shells of stable nuclei.
It has already been employed to study giant resonances of silicon and magnesium isotopes \cite{Peru:2008gd} and very recently the heavy
deformed $^{238}$U \cite{u8}. Exotic spherical $^{78}$Ni, $^{100}$Sn, $^{132}$Sn, and $^{208}$Pb  were previously studied
in the corresponding spherical HF+RPA calculation \cite{Peru:2005di}.

Some stellar phenomena, such as $r$-process nucleosynthesis, are particularly sensitive to the low-energy tails of dipole responses \cite{Goriely:1998}.
Theoretical microscopic studies of photoabsorption and radiative neutral capture cross sections \cite{Goriely:2002cx}
are essential in the regions of the nuclear chart where data are absent. All the models which pretend to be used in the whole
nuclear chart without changing any parameters must be tested against the available experimental data.
Also for this reason, after the calculations of \cite{Peru:2005di,Peru:2008gd,u8}  we focus here on  neon isotopes and $N=16$ isotones
since experimental data on $^{26}$Ne are available \cite{Gibelin:2008zz} and those on $^{24}$O will appear soon \cite{o24}.
Work on $^{68}$Ni is also in progress due to the experimental interest in the nature of the PDR \cite{Matea} in this nucleus \cite{Wieland:2009zz}.
In connection to this, an interesting article \cite{Carbone:2010az} relating the PDR to the nuclear symmetry energy \cite{Klimkiewicz:2007zz} and
to the isospin-dependent components of effective nuclear interactions recently appeared.

Our paper is organized as follows: We present our model in Sec. II and our results in Sec. III. Our work is summarized in Sec. IV.

\section{Model}
\label{model}

In this section, we briefly sketch our approach based on the QRPA on top of HFB calculations.
Details of the formalism,
as well as of $^{22-28}$Mg and  $^{26-30}$Si dipole responses can be found in \cite{Peru:2008gd}.
The HFB equations are solved in a finite harmonic oscillator (HO) basis.
As a consequence, the positive energy continuum is discretized.
We have checked the stability of the single-particle levels with the number of major HO shells.
Finally, according to \cite{Peru:2008gd}, we considered a model space including 9 HO major shells.
This is a large enough space for the neon isotopes and $N=16$ isotones considered here. 
To check it, we performed the calculation in a larger HO basis for
$^{26}$Ne. For 10 and 11 major shells, we obtain the same results as
for 9 major shells both for spurious and low-lying states. 
All HFB quasiparticle states are used to generate the 2-quasiparticle (2-qp) excitations.
This means that no cut in energy or in occupation probabilities is introduced.
As already emphasized, we use the same nucleon-nucleon effective force, the Gogny D1S \cite{gog1}, both for HFB and QRPA calculations in all particle-hole (ph),
particle-particle (pp), and hole-hole (hh) channels.
This is very important in order to avoid very dangerous inconsistencies. For this reason,
since the Coulomb exchange field is not taken into account in HFB,
the corresponding QRPA terms have been set to zero.

According to the symmetries imposed in the present axially-symmetric-deformed HFB calculations in even-even nuclei,
the projection $K$ of the angular momentum on the  symmetry axis and the parity $\pi$ are good quantum numbers.
Consequently, QRPA calculations can be performed separately in each $K^{\pi}$ block.
In an axially-symmetric-deformed nuclear system, the response function of  a given  $J^\pi$
contains different $K^\pi =0^\pi,\pm 1^\pi,...,\pm J^\pi$ components.
In spherical nuclei, all these components are degenerated in energy,
and then the response functions associated to any multi-polarity can be obtained from $K^\pi=0^{\pm}$ results only.
To solve the well-known (Q)RPA matrix equation
\begin{equation}\label{equaref}
\left(\begin{array}{cc} { A}& { B}\\{ B}&{ A} \end{array}  \right)
 \left(\begin{array}{c}{X^n}\\ {Y^n}\end{array}\right)
= \omega_n \left(\begin{array}{c}X^n\\-Y^n\end{array}\right),
\end{equation}
where  $\omega_n$ are the energies of the QRPA excited  states,
we use the same numerical procedure recently applied to study the giant resonances of the heavy deformed $^{238}$U \cite{u8}. It is based on a
massive parallel master-slave algorithm. For a single solution of Eq. (\ref{equaref}) the QRPA provides the set of amplitudes $X^n$ and $Y^n$
describing the wave function of the excited state $\vert n \rangle$ in terms of the two quasiparticle excitations.
Let us define $N_{\textrm{2qp}}$ as the number of all the possible 2-qp excitations for a given $K^{\pi}$ block. The well-known normalization
of the QRPA amplitude can then be written as
\begin{equation}
\sum_{\textrm{2qp=1}}^{N_{\textrm{2qp}}} [(X^{n}_{\textrm{2qp}})^2-(Y^{n}_{\textrm{2qp}})^2]=1.
\end{equation}
From this equation, it is easy to isolate the relative neutron and proton contributions by summing separately
over the two species. In the following, we indicate the corresponding percentage contribution with the notation $\%(\nu)$ and $\%(\pi)$.
In order to measure the degree of collectivity of a specific excited state,
we consider, inspired by \cite{Co':2009gi}, another index, called $N^{\star}$, which represents the number of states with
$[(X^{n}_{\textrm{2qp}})^2-(Y^{n}_{\textrm{2qp}})^2] \geq \frac{1}{N_{\textrm{2qp}}}$.
Also in this case we separate neutron ($N^{\star}_{\nu}$) and proton ($N^{\star}_{\pi}$) contributions.
Note that in Ref. \cite{Co':2009gi} RPA, instead of QRPA, calculations are performed. In this case, 
$N_{\textrm{2qp}}$ is replaced by $N_{\textrm{ph}}$, which represents the number of possible particle-hole excitations.
The $N^{\star}$ value in the RPA case (called here $N^{\star RPA} $) represents the number of states
with $[(X^{n}_{\textrm{ph}})^2-(Y^{n}_{\textrm{ph}})^2] \geq \frac{1}{N_{\textrm{ph}}}$.
In this case, when the excitation is produced by a single particle-hole state $N^{\star RPA}=1$
while in an ideal collective case all the particle-hole excitations contribute with the same statistical
weight so that $N^{\star RPA}=N^{\star RPA}_{\nu}+N^{\star RPA}_{\pi}=N_{\textrm{ph}}$.
In the QRPA, the situation is a little different because 2-qp excitations imply the inclusion of
hole-hole and particle-particle transitions.
Anyway, as the $N^{\star}$ values become greater, the response becomes more collective.
Obviously, the quantity $N_{\textrm{ph}}$, as well as $N_{\textrm{2qp}}$, is related to the size of the configuration space, 
which depends on the number of major oscillator shells one decides to consider and overall
on the symmetry scheme (spherical, axially-symmetric-deformed,...) of the calculation. For this reason, a direct comparison of our
results with the ones of Ref. \cite{Co':2009gi} is not possible. In Sec. \ref{n16isotones}, we perform a comparison
between this kind of analysis in RPA and QRPA schemes.
Note that even if for spherical nuclei the results obtained
for $K=0$ are strictly equivalent to the ones obtained for $K=1$, we choose to perform the analysis in the $K=0$ block.
In this way, the number of the possible 2-qp configurations for all the nuclei we consider is always the same,
precisely $N_{\textrm{2qp}}$=4832 (half obviously being neutron-neutron 2-qp excitations and half proton-proton 2-qp excitations).
Because we always consider the same number of possible 2-qp excitations, it is not necessary to introduce other indexes, such as the ratio
$N^{\star}/N_{\textrm{2qp}}$.

\section{Results}

\begin{figure}
\begin{center}
  \includegraphics[width=12cm,height=8cm]{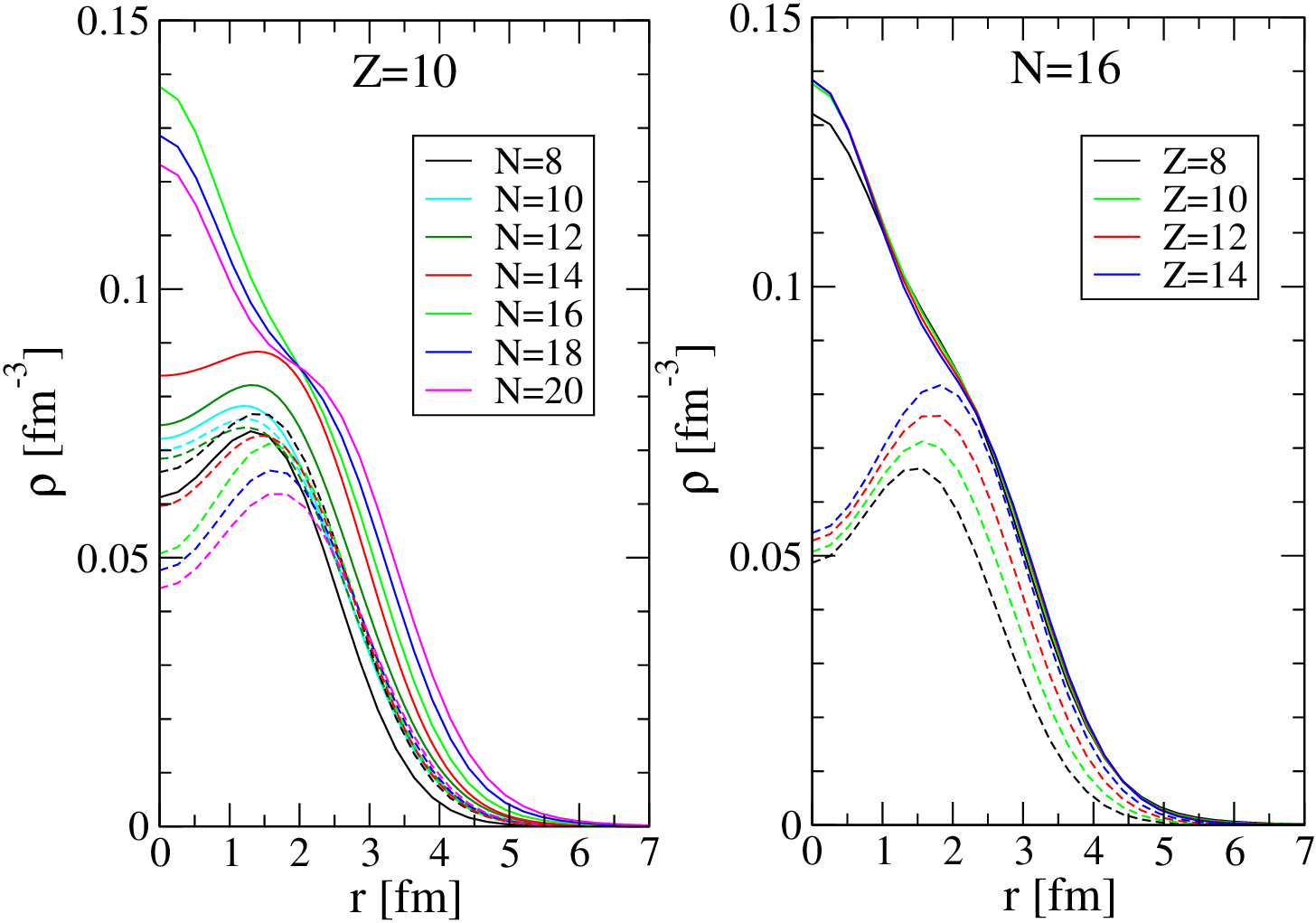}
\caption{(Color online) Neutron (solid lines) and proton (dashed lines) ground-state density profiles for neon isotopes (left panel)
and $N=16$ isotones (right panel) obtained through axially-deformed HFB calculation with D1S Gogny interaction.}
\label{fig_dens}
\end{center}
\vspace*{+0.8cm}
\end{figure}

The ground-state properties of the considered nuclei are obtained by solving HFB equations. Axially symmetric HFB+D1S results
for the whole nuclear chart can be found in \cite{Stephane}. For deformation parameters, neutrons and protons root-mean-square radii, pairing and Fermi
energies we refer the reader to this web site. In Fig. \ref{fig_dens}, we just plot the ground-state density profiles for the neon isotopes and $N=16$ isotones
evaluated in this approach with 9 HO major shells. In the case of deformed nuclei, the drawn density is the spherical projected one, expressed as 
a function of the $r$ spherical coordinate. For spherical nuclei the spherical radius $r$ coincides with the radial distance 
($r_{\perp}$) from the symmetry axis $z$ when the longitudinal position $r_z$ is $r_z=0$ or, equivalently, $r \equiv r_z$ 
when $r_{\perp}=0$. We observe in Fig. \ref{fig_dens} 
the formation and increase of a neutron skin on the nuclear surface when the neutron-proton ratio increases.
For $^{18}$Ne a proton skin appears.

\begin{figure}
\begin{center}
  \includegraphics[width=12cm,height=8cm]{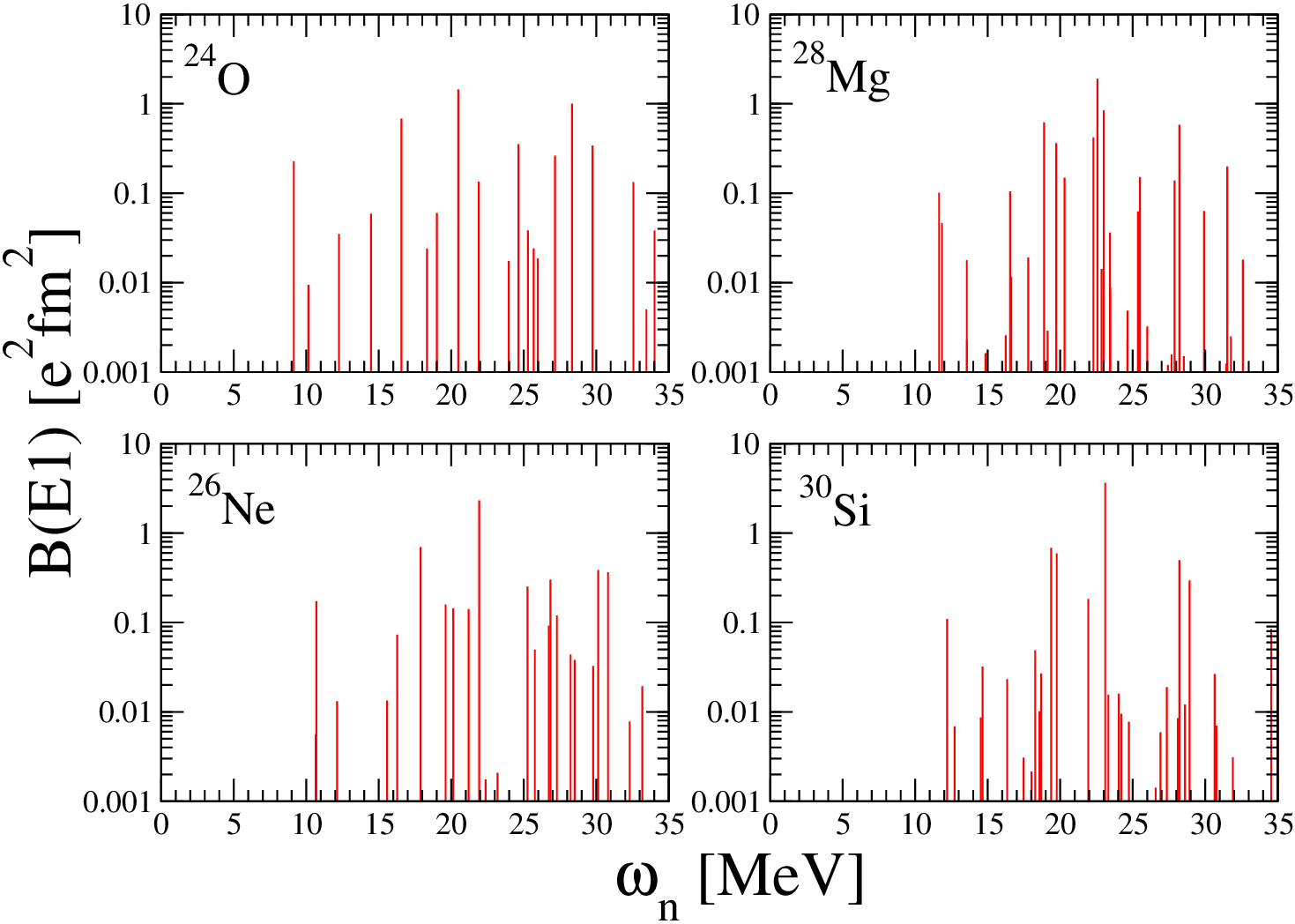}
\caption{(Color online) B(E1) distributions for $N=16$ isotones in logarithmic scale.}
\label{fig_be1_n16}
\end{center}
\vspace*{+0.8cm}
\end{figure}

\begin{figure}
\begin{center}
\includegraphics[width=12cm,height=8cm]{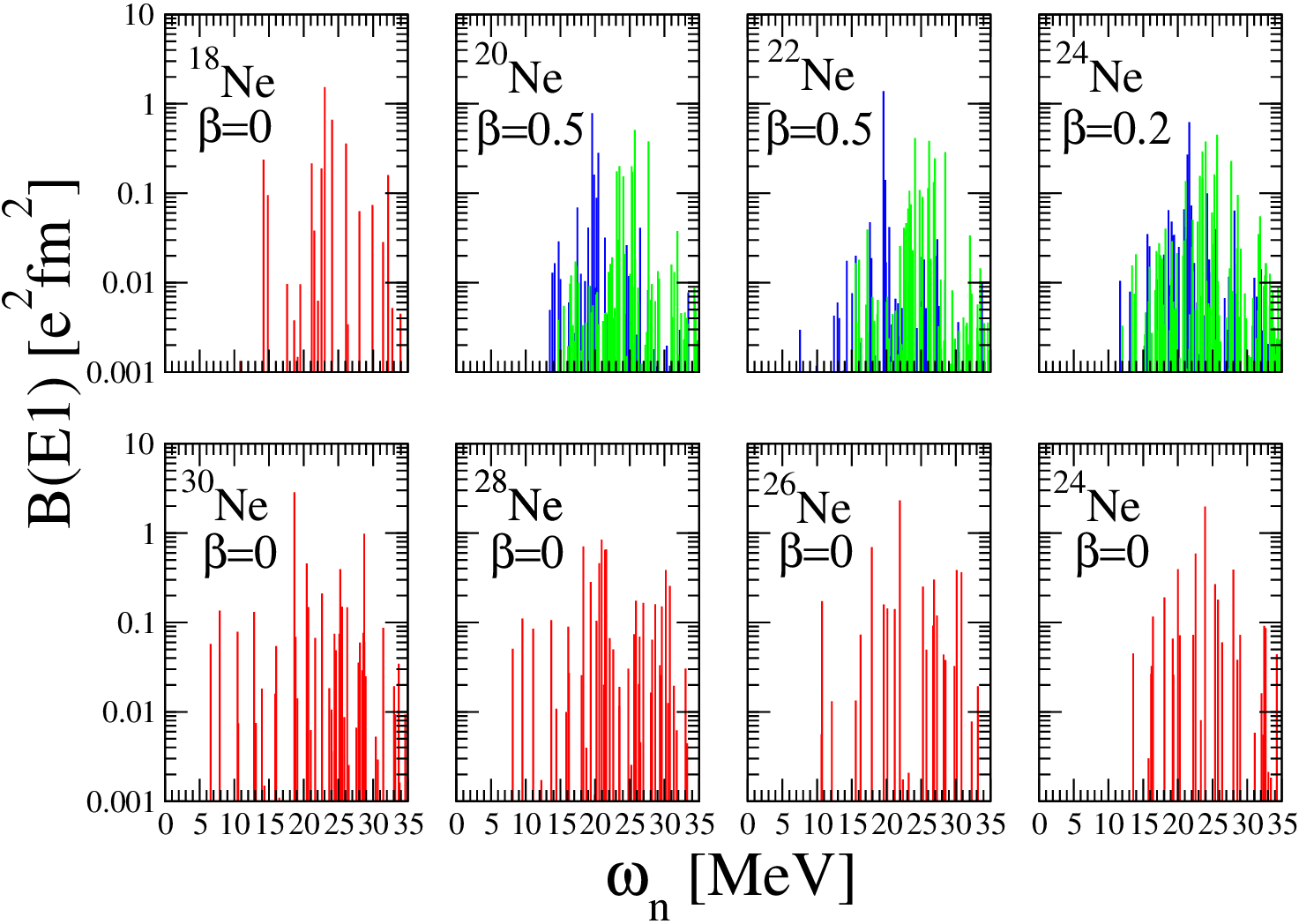}
\caption{(Color online) B(E1) distributions for neon isotopes in logarithmic scale.
For deformed isotopes $^{20,22,24}$Ne, the $K=0$ and $K=\pm 1$ components are drawn in black (blue online)
and gray (green online) respectively. In the spherical ($\beta=0$) case this splitting does not appear, and hence the
$K=0$ and $K=\pm 1$ coincide (red online).}
\label{fig_be1_z10}
\end{center}
\end{figure}

Figures \ref{fig_be1_n16} and \ref{fig_be1_z10} show the B(E1) distributions for $N=16$ isotones and neon isotopes respectively.
All the $N=16$ isotones we consider are spherical, according also to prediction of $N=16$ subshell closure from stability
to the neutron drip line \cite{Obertelli:2004qg}. Also, the analyzed neon isotopes, except
$^{20}$Ne ($\beta=0.5$), $^{22}$Ne ($\beta=0.5$) and $^{24}$Ne ($\beta=0.2$), turn out to be spherical.
We write in each panel of Fig. \ref{fig_be1_z10} the axial deformation parameter $\beta$ corresponding to the minimum of HFB potential energy.
For $^{24}$Ne the two minima at $\beta=0.2$ and $\beta=-0.2$ are quasidegenerated and the spherical ($\beta=0$) configuration
is just 100 keV less bound than the prolate one. For this reason, for $^{24}$Ne both the responses ($\beta=0.2$ and $\beta=0$) are given in Fig. \ref{fig_be1_z10}.
In the spherical symmetry case, the $K^{\pi}$=0$^{-}$ and $|K|^{\pi}$=1$^{-}$ states are degenerate. In deformed
$^{20}$Ne, $^{22}$Ne, and $^{24}$Ne nuclei the strength splits up into two components corresponding to two different angular momentum
projections $K$. As expected, since these nuclei are prolate, the GDR peak at low energy ($\omega_n \sim 20$ MeV) corresponds to the $K$=0 components.
The $|K|$=1 GDR strength is concentrated around 25 MeV.
For $^{20}$Ne and $^{22}$Ne nuclei, the $\omega_n < 14$ MeV excitations are negligible. In all the other cases one or some peaks appear in this low-energy region.
The giant resonance is located around 20-25 MeV.

\begin{figure}[t]
\begin{center}
\resizebox{.9\columnwidth}{!}{
\includegraphics{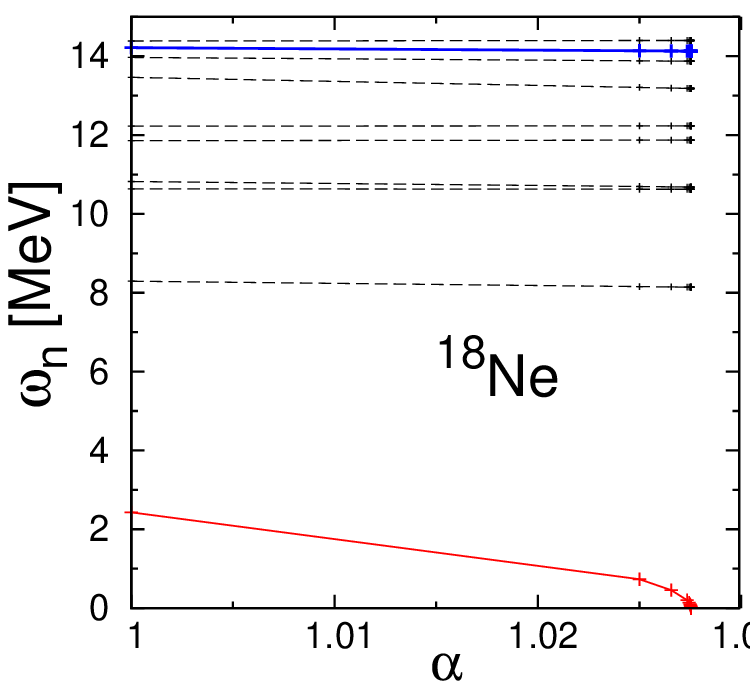}
\includegraphics{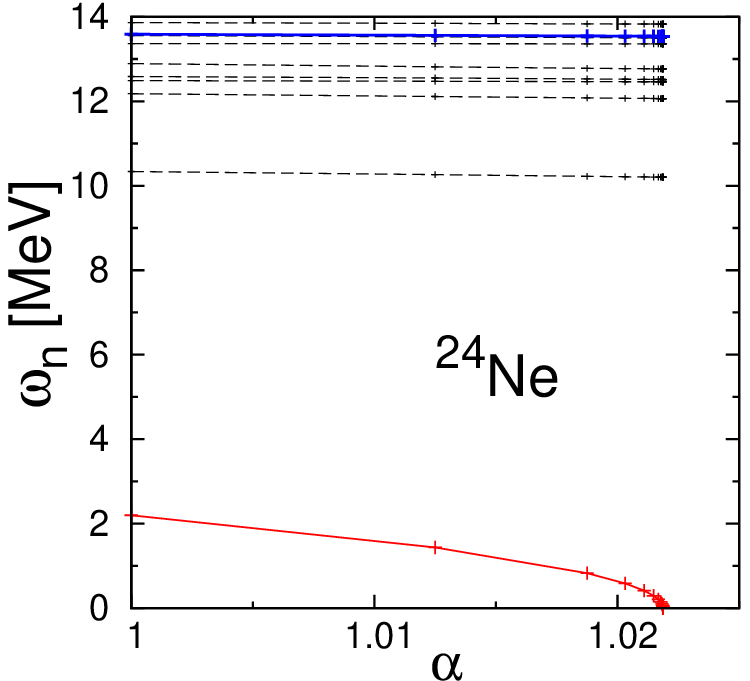}
\includegraphics{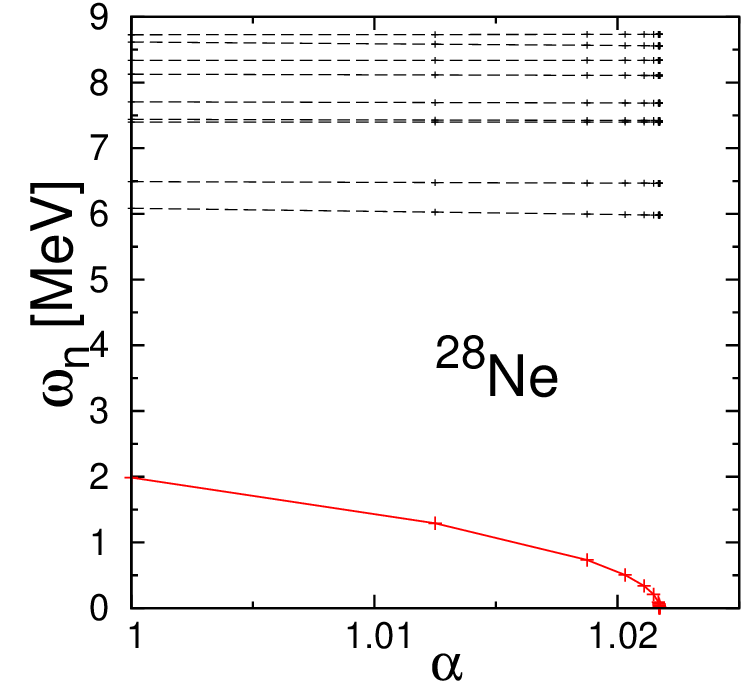}
\includegraphics{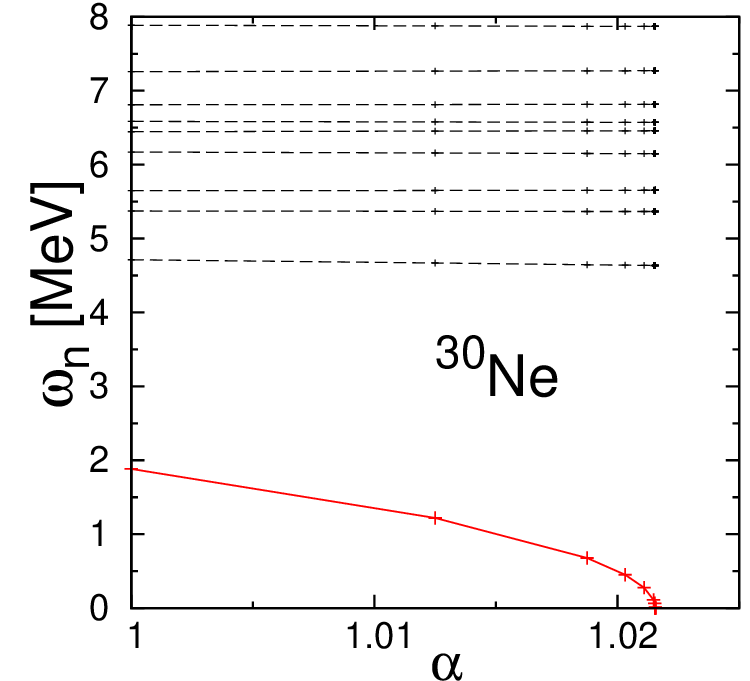}}
\resizebox{.9\columnwidth}{!}{
\includegraphics{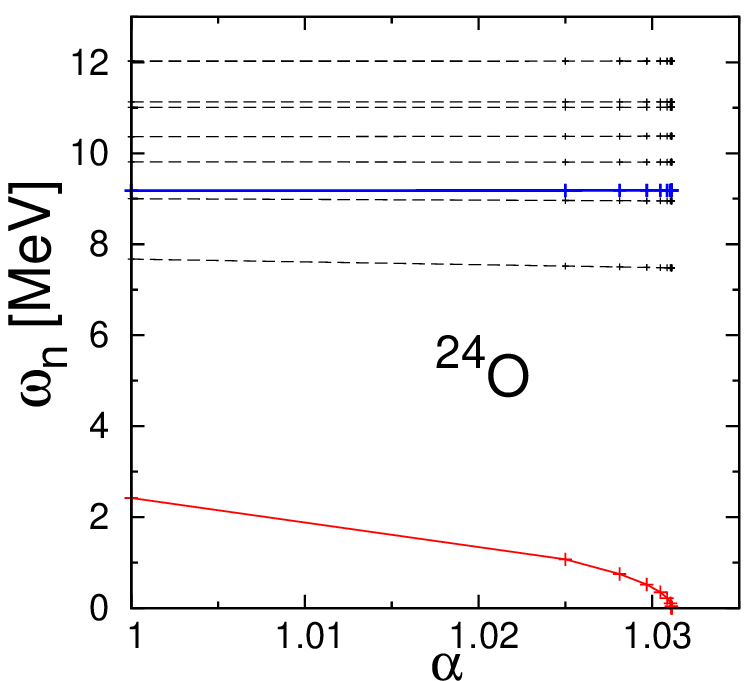}
\includegraphics{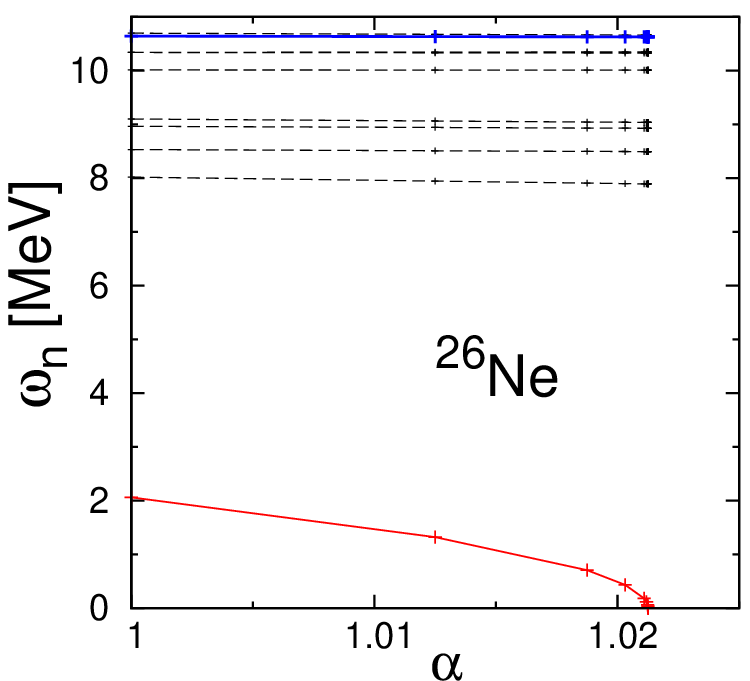}
\includegraphics{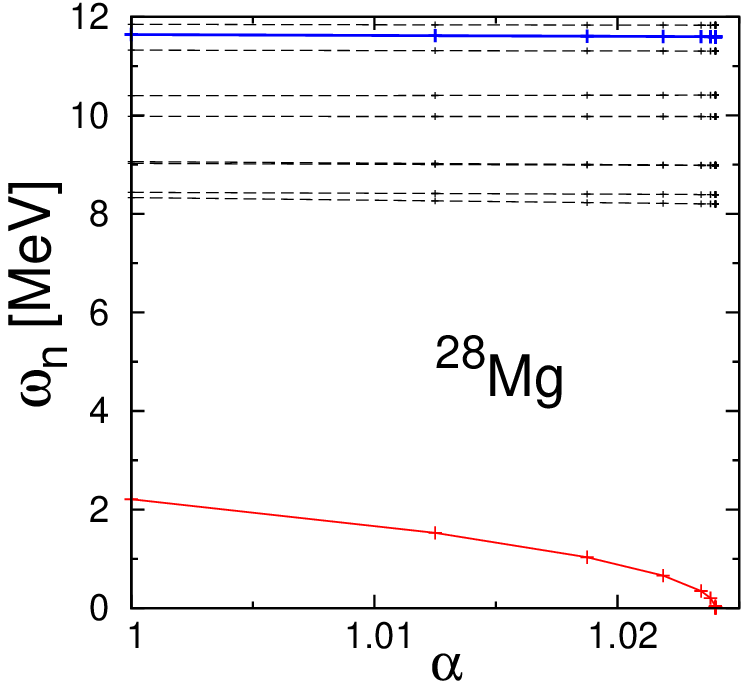}
\includegraphics{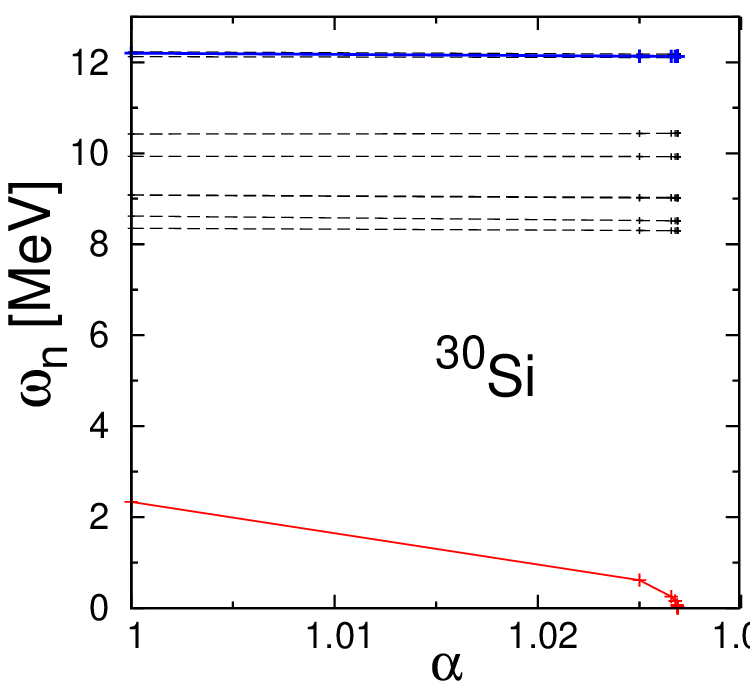}}
\end{center}
\caption{(Color online)
Evolution of the first ten $\omega_n$ eigenvalues with the $\alpha$ parameter for the neon isotopes and $N=16$ isotones, which present low-lying dipole excitations.
The lowest energy state is the spurious one. It is plotted with continuous line (red online) as well as the state corresponding to the 
analyzed excitation (blue online).
The remaining states are plotted with dashed black lines.}
\label{spurious}
\end{figure}

Each spectrum has been cleaned
by removing the corresponding translation spurious state. In order to identify it, a small renormalization factor $\alpha$ of the residual interaction has been introduced.
Figure \ref{spurious} shows the evolution of the first ten $\omega_n$ eigenvalues as a function of this $\alpha$ parameter. As clearly shown, only the spurious state
is affected by this factor while all the other states, including the ones corresponding to the low-energy peaks, stay quite constant. 
We checked also the stability of the B(E1) values with $\alpha$ in the case of $^{26}$Ne, and the strengths 
of all states but the spurious one remain always the same.
The spurious state, which turns out to be easily identifiable, is thus decoupled from the physical spectrum.

\subsection{$^{26}$Ne and $N=16$ isotones}
\label{n16isotones}

We start our analysis from $^{26}$Ne, which belongs both to $Z=10$ and $N=16$ chains. The low-energy dipole excitations of this nucleus
have been experimentally studied \cite{Gibelin:2008zz}. Several theoretical calculations on this nucleus have been already done \cite{Cao:2005bt,Pena:2007,Peru:2007,Yoshida:2008rw,Ebata:2010qr}.
In Table \ref{table1} we present a detailed analysis of the microscopic structure of the low-lying
dipole modes in $^{26}$Ne (and in $N=16$ isotones) considering the first low-energy state.
Instead to show the contribution of the main two quasiparticle excitations, we prefer to give the results in terms of the corresponding particle-hole (and particle-particle) transitions
identified by the usual spectroscopic notation. The energies corresponding to each single-particle transition for each nucleus are the values given between parentheses in Table \ref{table1}.
Focusing at first on $^{26}$Ne, it appears that the state at $\omega_n=10.7$ MeV is dominated by the two quasiparticle excitations corresponding to the $\nu$ 2s$_{1/2}$$\to$2p$_{3/2}$ transition.
In terms of percentages 
this contribution turns out to be of 67.6 \%, a value very close to the one of \cite{Yoshida:2008rw}.
According to \cite{Cao:2005bt}, the second contribution (9.5 \%) arises from the $\nu$ 2s$_{1/2}$$\to$2p$_{1/2}$ excitation.
Two-quasiparticle proton excitations are 8.1  \%. An important feature of our results is an appreciable contribution (8.7 \%) of the $\nu$ 1d$_{5/2}$$\to$1f$_{7/2}$ configuration.
The experimental results of \cite{Gibelin:2008zz} seem to suggest that the low-lying dipole excitations around 6-10 MeV involve more transitions with respect
to those only characterized by 2s$_{1/2}^{-1}$ states. 
As stressed in \cite{Gibelin:2008zz} theoretical private calculations in the approach of \cite{Pena:2007} predict a nearly equal weight of 
$\nu$ 2s$_{1/2}$$\to$2p$_{1/2}$ and $\nu$ 1d$_{5/2}$$\to$1f$_{7/2}$ transitions in the low lying $K^{\pi}=1^-$, which turns out to be dominant in this model. 
In connection with the experimental results of \cite{Gibelin:2008zz}, also the structure of the second main peak, at $\omega_n=17.9$ MeV is interesting.
It is mainly generated by the  $\nu$ 1p$_{1/2}$$\to$1d$_{3/2}$ (49.3  \%), $\nu$ 1d$_{5/2}$$\to$1f$_{7/2}$ (24.4  \%) and $\nu$ 1d$_{5/2}$$\to$2p$_{3/2}$ (14.42 MeV) (13.1  \%) transitions.

\begin{figure}
\begin{center}
  \includegraphics[width=12cm,height=8cm]{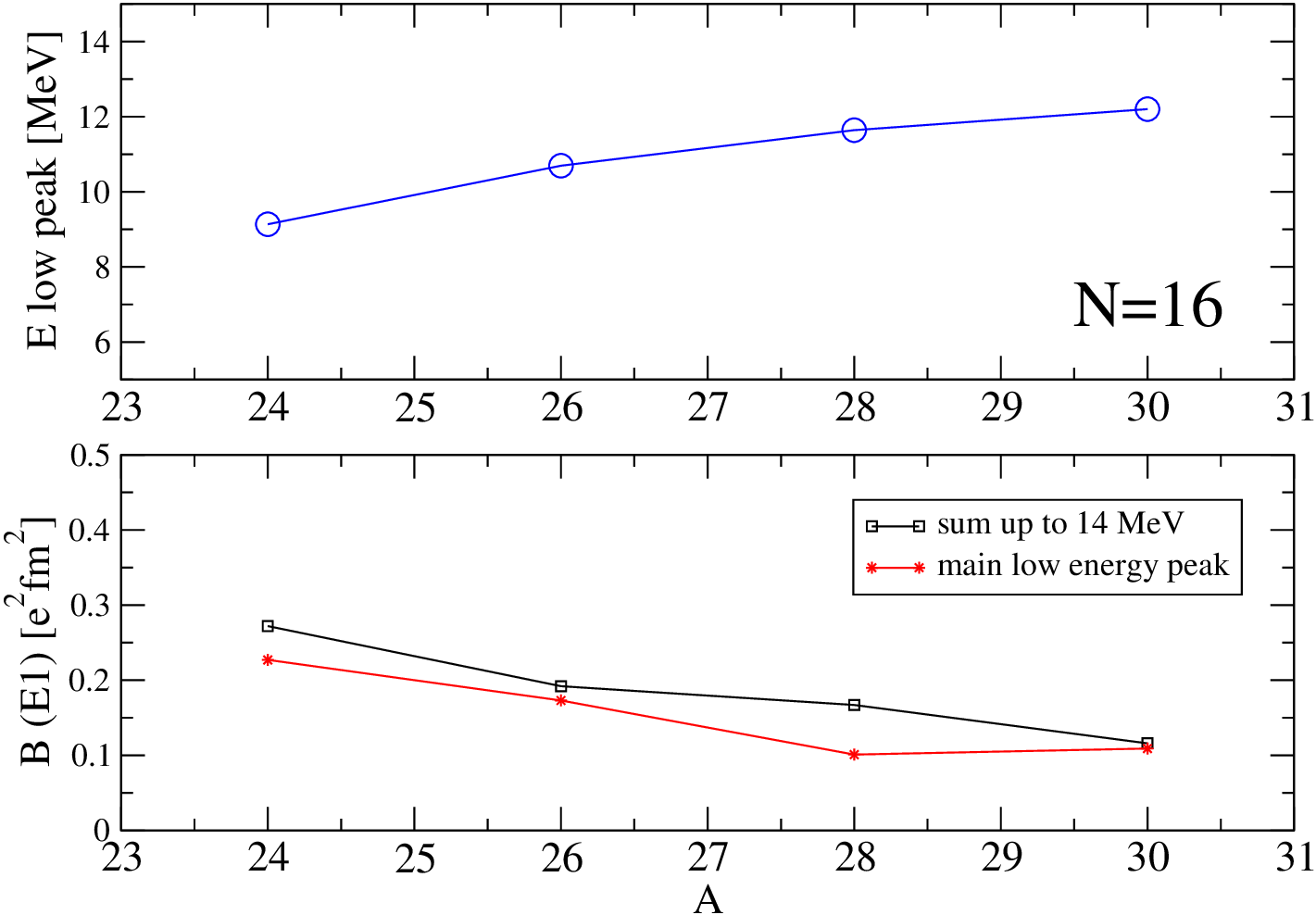}
\caption{(Color online) Energy of the low-lying peak (upper panel) and B(E1) transition strengths (lower panel) for the $N=16$ isotones as functions of mass number.}
\label{fig_N16_vs_A}
\end{center}
\vspace*{+0.8cm}
\end{figure}

Concerning the $N=16$ isotones, the other nuclei we consider here are $^{24}$O, $^{28}$Mg, and $^{30}$Si.
While the low-energy peak is expected in $^{24}$O, one can observe in Fig. \ref{fig_be1_n16} that an equivalent structure
appears at $\omega_n \simeq12$ MeV also for $^{28}$Mg and $^{30}$Si in spite of their small neutron-proton asymmetry.
As expected and as shown in Fig. \ref{fig_N16_vs_A} the centroid energy of the low-lying E1 states increases with the decrease
of neutron excess and the B(E1) value of the first peak is suppressed. On the other hand, the total strength up to 14 MeV decreases
very slowly from $^{26}$Ne to $^{30}$Si.

Turning to the microscopic analysis of Table \ref{table1}, one can observe that the structure of the
first peak in $^{24}$O is quite similar to the one of $^{26}$Ne except for the absence
in $^{24}$O of the $\nu$ 1d$_{3/2}$$\to$2p$_{3/2}$  contribution, as expected for a particle-particle configuration in a closed-shell nucleus.
This particle-particle transition turns out to be very important in $^{28}$Mg,
being of the same order of magnitude as the main particle-hole transition $\nu$ 2s$_{1/2}$$\to$2p$_{3/2}$. Finally, in  $^{30}$Si the contribution of proton excitations 
becomes appreciable, being of the order of 30\%. In the three other cases, the neutron contribution to the low energy peak is more than 90\%. 
Even if a detailed study of the GDR region is beyond the aim of this paper we consider here and in the following the behavior of the 
main GDR peak in order to compare it with the one of low-lying states. 
For this reason in Table \ref{table1} the total proton and neutron contributions to the main GDR peak are reported. The difference 
with respect to the corresponding first peak clearly appears. 
It is interesting to observe
that even in the GDR peak the neutron contribution is larger than the expected $N/A$ value in the cases of $^{24}$O and $^{28}$Mg nuclei. 
Note always from Table \ref{table1} that if instead of treating only the main GDR peak we consider 
the total proton and neutron contribution arising 
from all the states with 15 MeV $\leq \omega_n \leq$ 30 MeV and B(E1) $\geq$ 0.5 e$^2$ fm$^2$, the result is quite similar.

Further remarks about the collective behavior of the analyzed dipole excitations in these $N=16$ isotones arise from the calculation of the $N^{\star}$,
$N^{\star}_{\nu}$ and $N^{\star}_{\pi}$ indexes defined in Sec. \ref{model} and reported in Table \ref{table_N16_isoton_Nstar_index} together with the isospin percent contributions
$\%(\nu)$ and $\%(\pi)$.
Because we start the analysis from $^{24}$O, some comments are needed. The little values of $N^{\star}$ numbers (even in the GDR case) with respect to $N_{\textrm{2qp}}$ must not be surprising.
This apparent disproportion is related to the huge number $N_{\textrm{2qp}}=4832$, which is due to the choice of performing
calculation in QRPA, including in this way not only the particle-hole transitions but also the particle-particle and the hole-hole ones
without any cut in excitation energy. It is important to observe that in the corresponding
spherical RPA calculation (which for $^{24}$O is strictly equivalent to the QRPA one) the number of possible particle-hole excitation is largely reduced ($N_{\textrm{2qp}}=4832 \to N_{\textrm{ph}}=64$),
increasing the relative importance of $N^{\star}$ indexes, which in this case become  $N^{\star RPA}_{\nu}$=3 and $N^{\star RPA}_{\pi}$=2 at $\omega_n$=9.1 MeV and
$N^{\star RPA}_{\nu}$=7 and $N^{\star RPA}_{\pi}$=3 at $\omega_n$=20.5 MeV.

Coming back to the analysis of the QRPA results shown in Table \ref{table_N16_isoton_Nstar_index}, one can observe that
the number of state which contribute with a weight greater than $\frac{1}{N_{\textrm{2qp}}}$, as defined in Sec. \ref{model}, lightly increases from
$^{24}$O to $^{30}$Si both for low-energy mode and for the GDR. The $N^{\star}_{\nu}$ index generally decreases with $Z$ while $N^{\star}_{\pi}$ increases.
The number of configurations which contribute to the low energy mode is about the 60\% of the configurations involved in the GDR. This value reduces to about 40\% just in the
$^{28}$Mg case. This apparently less collective behavior of the 11.6 MeV peak is compensated by the appearance of another quite close peak at 11.8 MeV, which is dominated by
the $\nu$ 1d$_{3/2}$$\to$2p$_{3/2}$ (62\%) and the $\nu$ 2s$_{1/2}$$\to$2p$_{3/2}$ (24\%) transitions.

\begin{figure}
\begin{center}
  \includegraphics[width=12cm,height=8cm]{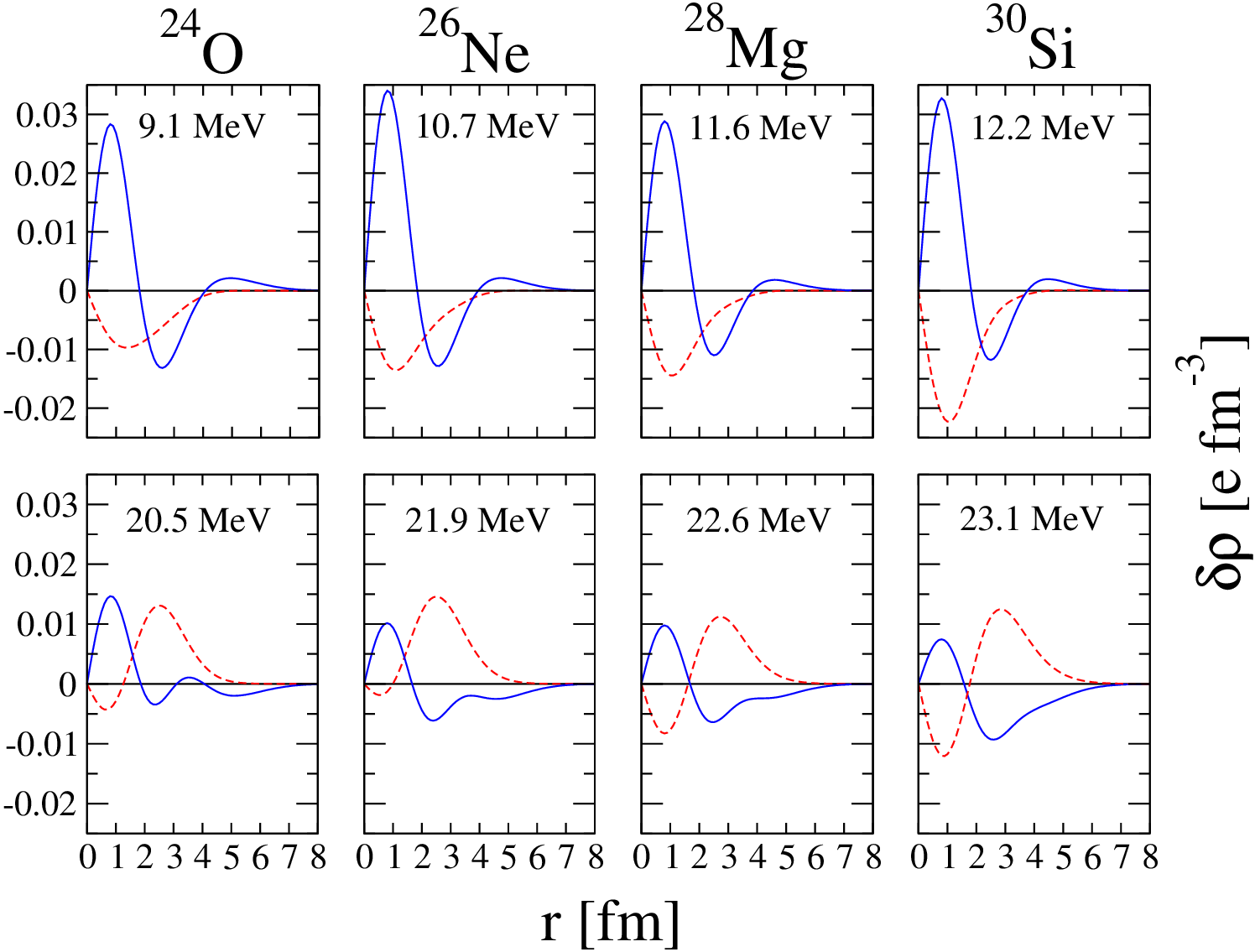}
\caption{(Color online) Neutron (full lines) and proton (dashed lines) transition densities for the
$^{24}$O, $^{26}$Ne, $^{28}$Mg and $^{30}$Si nuclei.
The corresponding excitation energies are given in the panels.}
\label{fig_dens_trans_n16}
\end{center}
\vspace*{+0.8cm}
\end{figure}

\begin{figure}
\begin{center}
  \includegraphics[width=12cm,height=8cm]{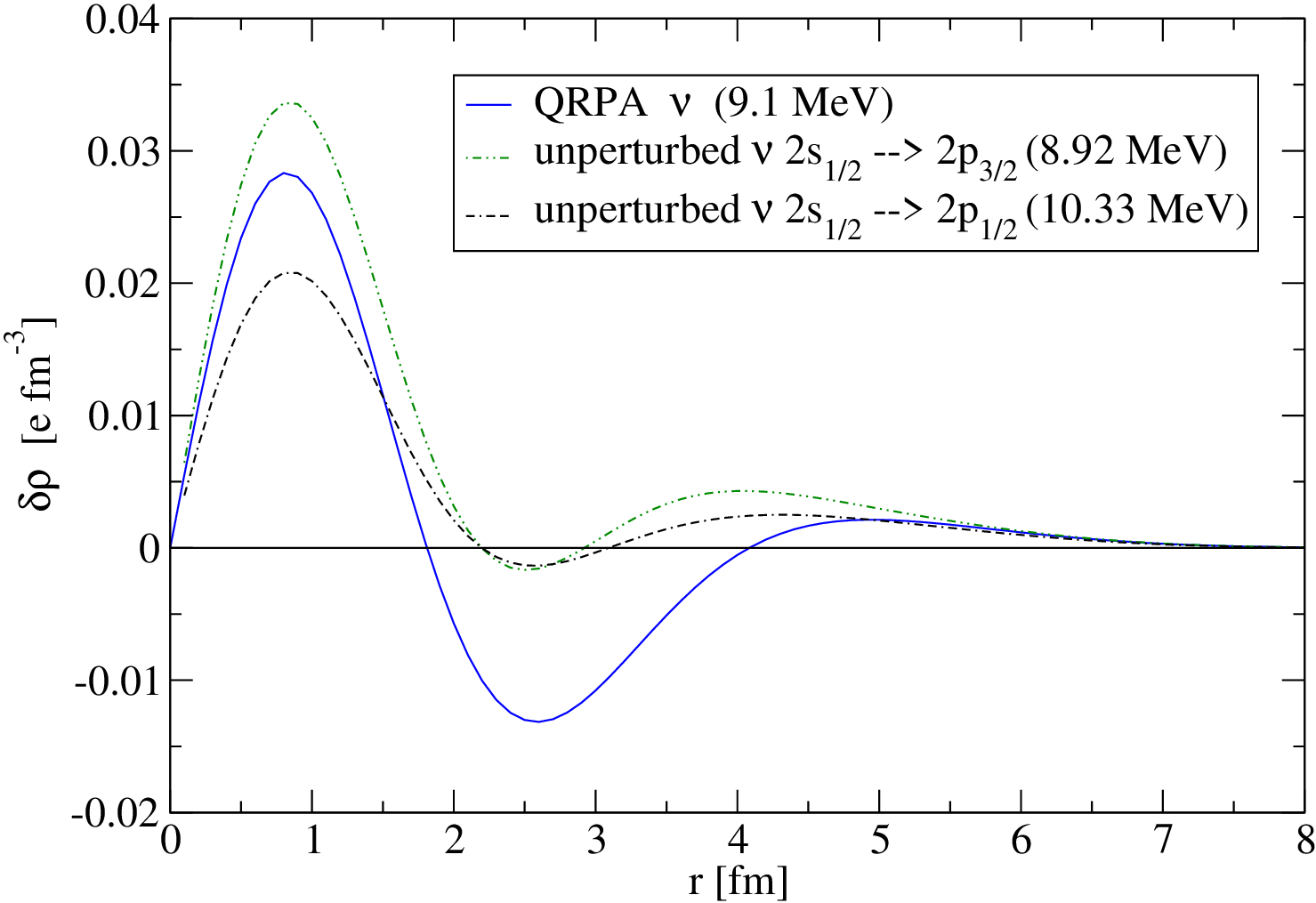}
\caption{(Color online) $^{24}$O neutron transition density for the first low-energy peak at 9.1 MeV calculated in QRPA (full line) 
compared to the unperturbed transition densities of $\nu$ 2s$_{1/2}$$\to$2p$_{3/2}$ (dot-dot-dashed line) 
and  $\nu$ 2s$_{1/2}$$\to$2p$_{1/2}$ (dot-dash-dashed line) excitations.}
\label{fig_dens_trans_o24}
\end{center}
\vspace*{+0.8cm}
\end{figure}

In Fig. \ref{fig_dens_trans_n16} we display the neutron and proton transition densities $\delta \rho$ for $^{26}$Ne and for the other
$N=16$ isotones for the low-energy and the main GDR peaks. Note that since the nuclei we considered turn out to be spherical, 
even if we performed our calculation in cylindrical coordinates,
we can and prefer to present the transition densities as a function of $r$ instead that as a function of $r_z$ and $r_{\perp}$. 
The lower panels of Fig.\ref{fig_dens_trans_n16} refer to the main GDR peak. 
They display for all the $N=16$ isotones a radial dependence that is
characteristic for the isovector GDR: The proton and
neutron densities oscillate with opposite phases.
The low-energy case, shown in the upper panel, is more complex but is quite similar in all the $N=16$ isotones.
It presents an isovector character in the center of the nucleus (for $r \leq 2$ fm) and an isoscalar one in the surface region ($r \simeq$ 3 fm).
For  $r \geq 4$ fm one can observe an oscillating finite contribution
of the neutrons and a negligible one of the protons. The spatially extended structure of the 2s$_{1/2}$
state is responsible for the shape of the neutron transition density. In order to illustrate this point we show in 
Fig. \ref{fig_dens_trans_o24} the $^{24}$O neutron transition density for the state at 9.1 MeV, as already plotted 
in the first panel of Fig. \ref{fig_dens_trans_n16}, and the unperturbed transition densities for 
$\nu$ 2s$_{1/2}$$\to$2p$_{3/2}$ (8.92 MeV) and $\nu$ 2s$_{1/2}$$\to$2p$_{1/2}$ (10.33 MeV) excitations. It clearly appears 
that the behavior at the center of the nucleus is quite similar, reflecting the crucial role of the  $\nu$ 2s$_{1/2}$ state, 
while in the surface region the two unperturbed transition densities differ from the one obtained in QRPA, reflecting the important 
role of correlations in this region. For  $r \geq 5$ fm, the three transition densities turn to be similar.

The result obtained for the transition densities of the low-energy excitations of $N=16$ isotones does not coincide with the usual isoscalar representation of the PDR obtained in many theoretical calculations but for heavier nuclei
\cite{Paar:2007bk}. 
Transition densities of the low-lying dipole states in $^{26}$Ne can be found in \cite{Yoshida:2008rw} and in \cite{Pena:2007}, 
while results for $^{24}$O are presented in \cite{Matsuo:2004pr} and \cite{Co':2009gi} where a phenomenological RPA is used. 
In \cite{Matsuo:2004pr}, the isoscalar character 
of the excitation inside the nucleus clearly appears while the isovector behavior of the analyzed low energy excitation 
(however not the lowest one) in \cite{Co':2009gi} is interpreted as the tail part of the GDR. 
At the moment, the results obtained for the transition densities of the low-lying dipole state in these nuclei 
seem to be model dependent, and the problem remains open.

To summarize, the low-lying dipole state of $N=16$ isotones possesses a small but finite collective nature.
While the strength of this state increases with the neutron-to-proton ratio, the degree of collectivity stays quite constant.
These considerations, as well as the study of the transition densities, suggest that some interesting low-energy properties can be experimentally obtained
also from the less exotic $^{28}$Mg and $^{30}$Si nuclei.

\subsection{Neon isotopes}

\begin{figure}
\begin{center}
  \includegraphics[width=12cm,height=8cm]{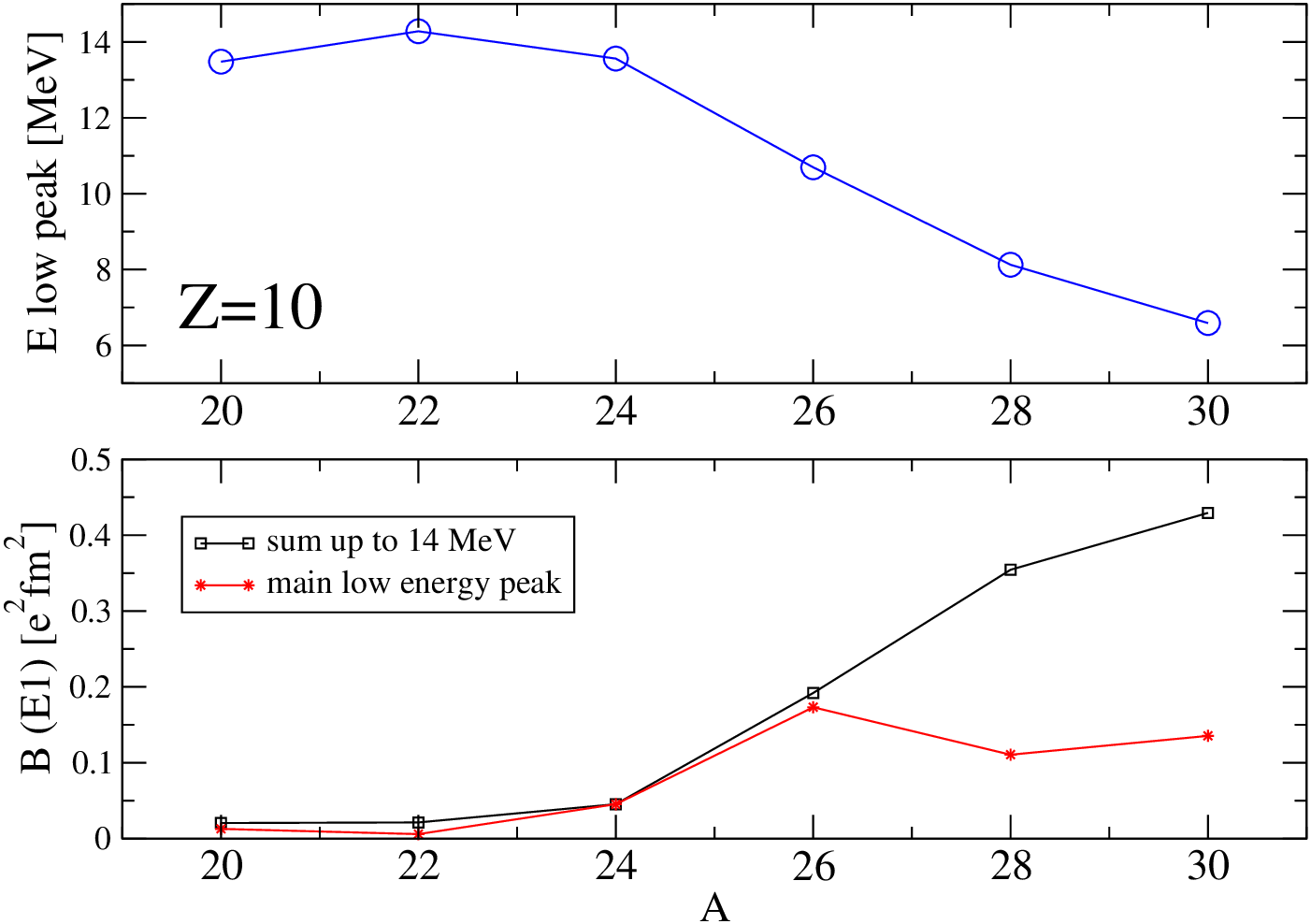}
\caption{(Color online) Energy of the first low-energy peak with B(E1)$\geq$ 0.01 e$^2$ fm$^2$ (upper panel) and B(E1) transition strengths (lower panel) for the 
neon isotopes as functions of mass number.}
\label{fig_ne_isot_vs_A}
\end{center}
\vspace*{+0.8cm}
\end{figure}

The analysis of neon isotopes proceeds along the same lines. 
Figure \ref{fig_ne_isot_vs_A} is the equivalent of Fig. \ref{fig_N16_vs_A} but for the $Z=10$ instead of $N=16$ chain. Also in this case,
the energy of the low-lying peak decreases with the isospin asymmetry and the B(E1) strength increases. As shown in the lower panel of
Fig.\ref{fig_ne_isot_vs_A} by comparing the B(E1) value of the low-energy peak with the value obtained summing the B(E1) up to 14 MeV
the fragmentation increases with the number of neutrons. It merges also from Fig. \ref{fig_be1_z10}.

\begin{figure}
\begin{center}
  \includegraphics[width=12cm,height=8cm]{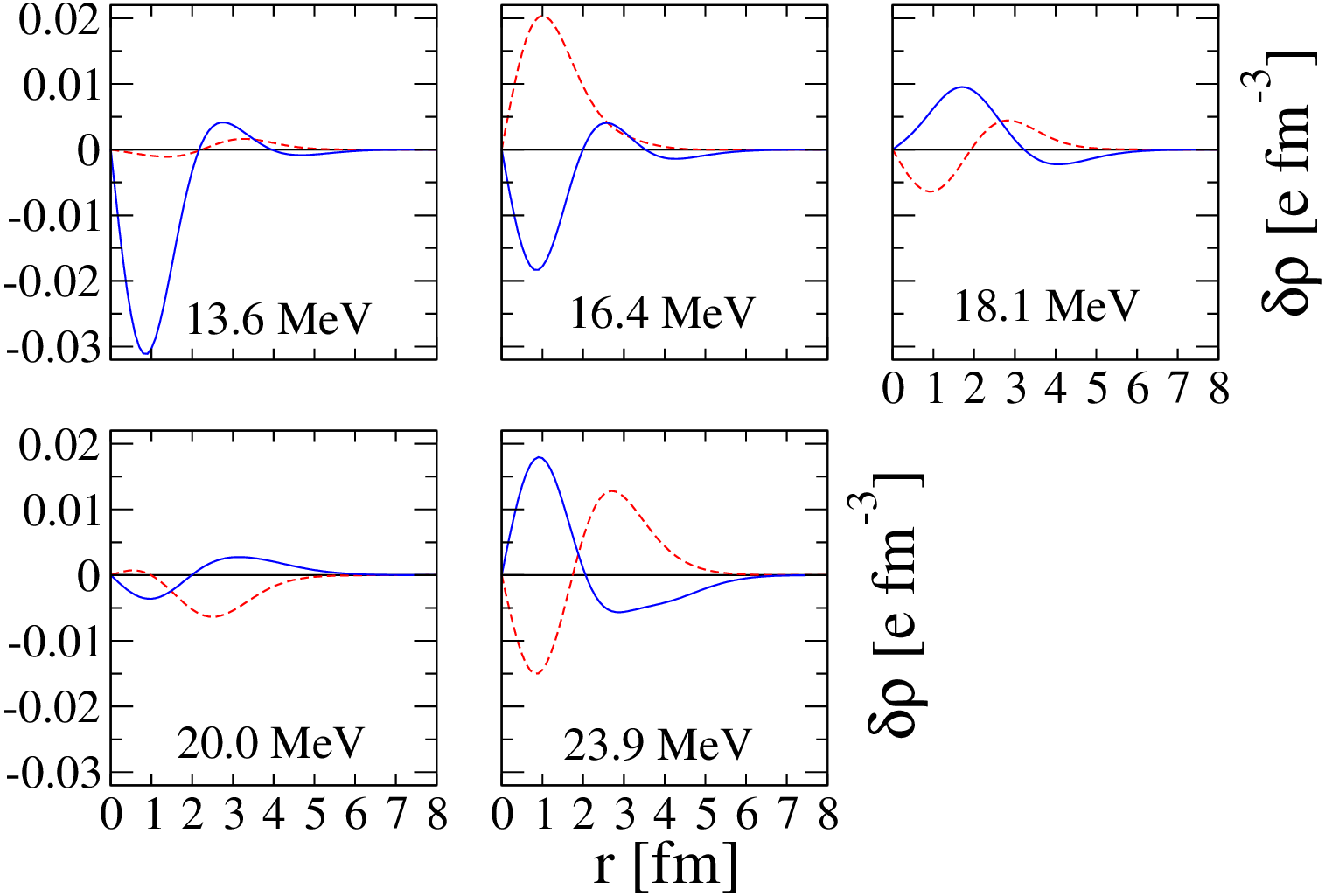}
\caption{(Color online) Neutron (full lines) and proton (dashed lines) transition densities for $^{24}$Ne.
The corresponding excitation energies are given in the panels.}
\label{fig_dens_trans_ne24}
\end{center}
\vspace*{+0.8cm}
\end{figure}

The lower mass isotope, which presents a non-negligible strength, is $^{24}$Ne, as recently found also in \cite{Ebata:2010qr}. 
According to the HFB ground-state calculation
\cite{Stephane}, this nucleus turns out to be lightly deformed. Nevertheless, owing to the soft deformation with prolate ($\beta=0.2$) and oblate
($\beta=-0.2$) configurations mostly degenerate, we prefer to consider $^{24}$Ne as spherical in our microscopic analysis. The comparison with other neon isotopes
becomes in this way more easy and more consistent. Anyway, in Fig. \ref{fig_be1_z10} we reported the B(E1) results for $^{24}$Ne both in $\beta=0.2$ and $\beta=0$ cases.
The deformed response is obviously more fragmented with the splitting of $K=0$ and $K=1$ components but the correspondences
of low-energy and GDR excitations in $\beta=0.2$ and $\beta=0$ cases are evident.
For example, focusing on the little low-energy excitation, one can observe that the single peak at $\omega_n$=13.6 MeV with B(E1)=4.5 10$^{-2}$ e$^2$ fm$^2$ in the spherical case
is split, in the prolate case, in one peak for the $K=0$ component at
$\omega_n$=13.1 MeV with B(E1)=0.8 10$^{-2}$ e$^2$ fm$^2$ and in three $K=1$ peaks
at $\omega_n$=13.4 MeV,  $\omega_n$=13.8 MeV and $\omega_n$=13.9 MeV with
B(E1)=1.6 10$^{-2}$ e$^2$ fm$^2$, B(E1)=0.7 10$^{-2}$ e$^2$ fm$^2$ and B(E1)=2.1 10$^{-2}$ e$^2$ fm$^2$, respectively.
Turning to the microscopic analysis in the spherical case,
the peak at 13.6 MeV is mostly due (65.4\%), as in $^{26}$Ne, to
$\nu$ 2s$_{1/2}$$\to$2p$_{3/2}$ (15.03 MeV) transition. In this case, the $\nu$ 2s$_{1/2}$$\to$2p$_{1/2}$ (15.93 MeV) transition turns out to be more
relevant (25.7\%) with respect to $^{26}$Ne. Two other transitions, $\nu$ 1d$_{3/2}$$\to$2p$_{1/2}$ (19.78 MeV) and $\pi$ 1p$_{1/2}$$\to$2s$_{1/2}$ (16.14 MeV), are of the order of 2\%.
Each of the remaining ones contributes individually less than 1\%. The transition densities of this state, shown in Fig. \ref{fig_dens_trans_ne24}, are quite interesting
revealing a neutron fluctuation larger than the proton one. The very small proton fluctuation is anyway in phase with
the neutron one, reflecting the isoscalar nature of this resonance. In contrast, the transition densities corresponding to all the other remarkable peaks shown an evident
isovector behavior typical of the GDR. Furthermore, as deduced from $N^{\star}_{\nu}$ and $N^{\star}_{\pi}$ indexes reported in Table \ref{table_ne_isotop_Nstar_index},
the excitation at 13.6 MeV involves principally neutron configurations and is little more collective than the successive one at 16.4 MeV.

\begin{figure}
\begin{center}
  \includegraphics[width=12cm,height=8cm]{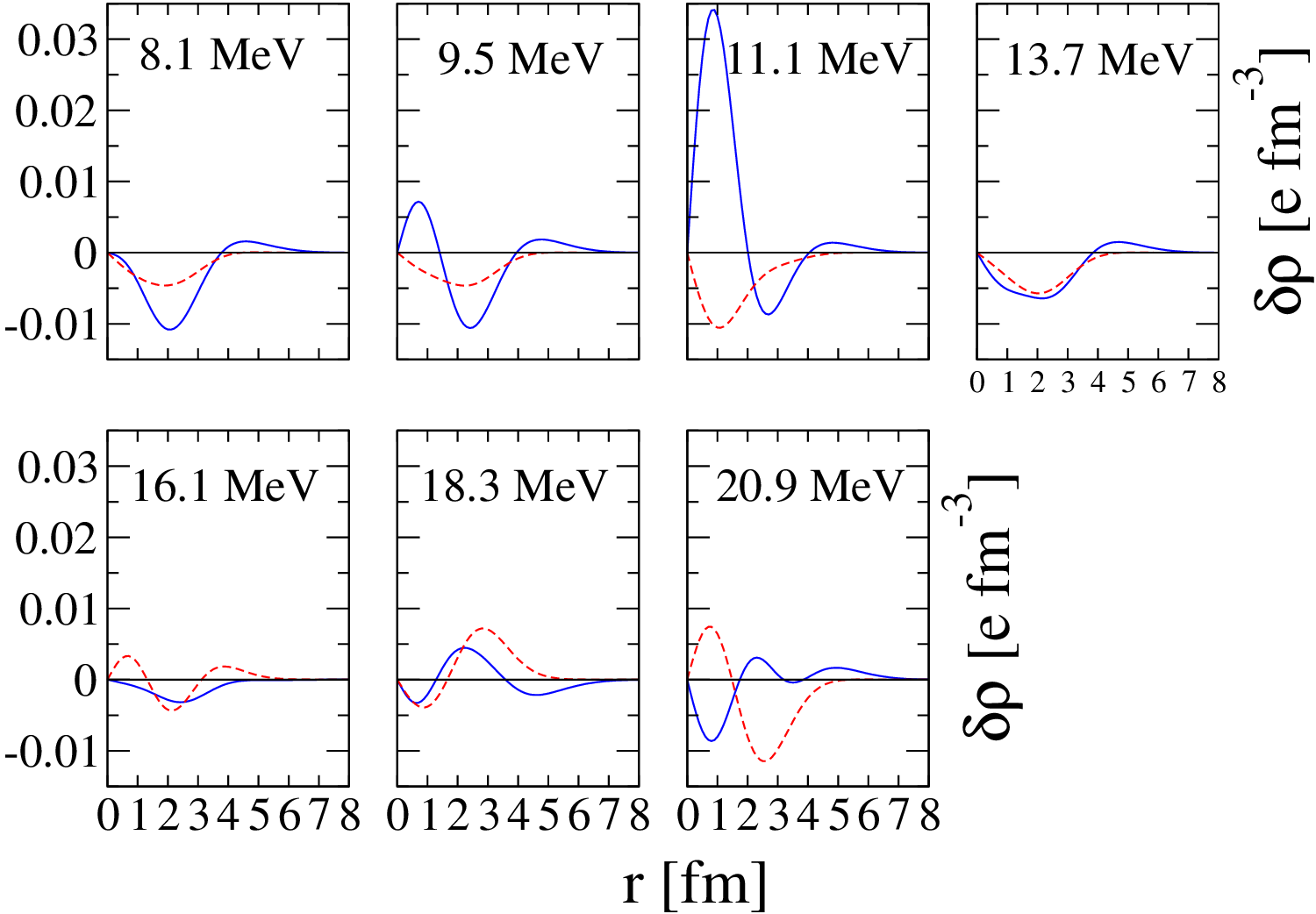}
\caption{(Color online) Neutron (full lines) and proton (dashed lines) transition densities for $^{28}$Ne.
The corresponding excitation energies are given in the panels.}
\label{fig_dens_trans_ne28}
\end{center}
\vspace*{+0.8cm}
\end{figure}

Since we have already discussed in the previous subsection $^{26}$Ne, we turn now to $^{28}$Ne. In this case, the dipole strength distribution is more fragmented.
The first two peaks are essentially of single-particle nature. They are related to the transitions from the less bounded 1d$_{3/2}$ neutron state:
$\nu$ 1d$_{3/2}$$\to$2p$_{3/2}$ (8.26 MeV) (90.4\%) for the 8.1 MeV peak and
$\nu$ 1d$_{3/2}$$\to$2p$_{1/2}$ (10.00 MeV) (83.1\%) for the second one at 9.5 MeV.
The corresponding  $N^{\star}_{\nu}$ and $N^{\star}_{\pi}$ indexes shown in Table \ref{table_ne_isotop_Nstar_index}
are lower with respect to the ones related to higher energy excitations.
The state at 11.1 MeV corresponds to the low energy peak of $^{26}$Ne. The dominant component is the same, that is, $\nu$ 2s$_{1/2}$$\to$2p$_{3/2}$ (10.82 MeV) (67.2\%).
The degree of collectivity is very similar, according to Table \ref{table_ne_isotop_Nstar_index},
as well as the neutron and proton transition densities illustrated in Fig. \ref{fig_dens_trans_ne28}.
The successive peak at 13.7 MeV is really interesting. As reported in Table \ref{table_ne_isotop_Nstar_index}, the number of 2-qp neutron excitations related
to this state is appreciably greater than the one of the lower states and even of the successive high-energy state.
In this 13.7 MeV peak, the transition from 1d$_{3/2}$ neutron state turns to play a major role: 68.8 \% from $\nu$ 1d$_{3/2}$$\to$1f$_{5/2}$ (14.09 MeV).
The prominent role of transitions from the weakly bound 1d$_{3/2}$ neutron state is reflected in neutron densities, no more peaked at the interior of the nucleus
as for the transitions involving the $\nu$ 2s$_{1/2}$ state. The isoscalar behavior of the state dominated by these $\nu$ 1d$_{3/2}$ transitions is also evident.

\begin{figure}
\begin{center}
  \includegraphics[width=12cm,height=8cm]{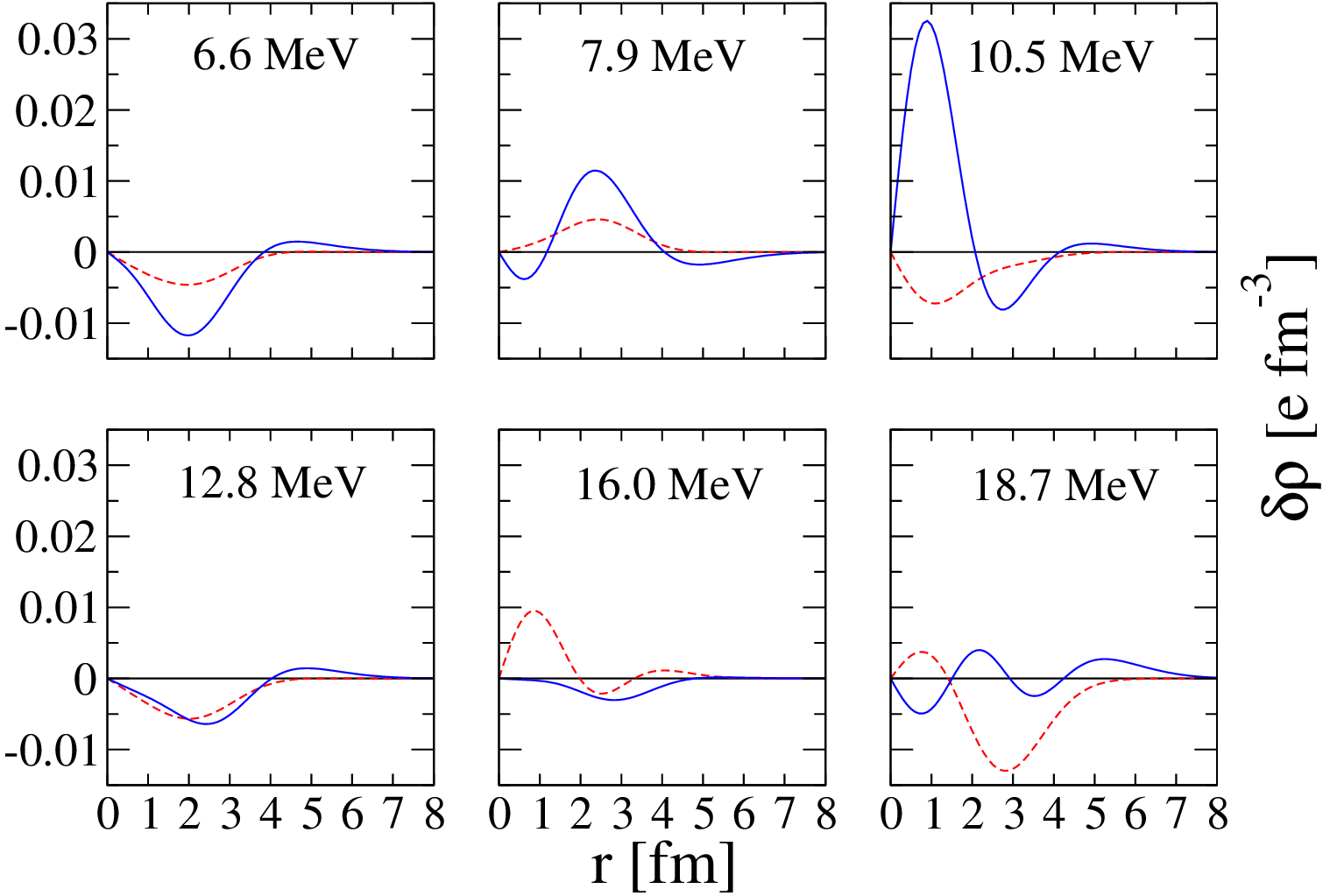}
\caption{(Color online) Neutron (full lines) and proton (dashed lines) transition densities for $^{30}$Ne.
The corresponding excitation energies are given in the panels.}
\label{fig_dens_trans_ne30}
\end{center}
\vspace*{+0.8cm}
\end{figure}

Results obtained for $^{30}$Ne are very similar to those for $^{28}$Ne concerning both the microscopical analysis and the transition densities
shown in Fig. \ref{fig_dens_trans_ne30}.
The first two peaks are still less collective: 93.5\% from $\nu$ 1d$_{3/2}$$\to$2p$_{3/2}$ (6.57 MeV) in the first peak at 6.6 MeV and 89.9\% from
$\nu$ 1d$_{3/2}$$\to$2p$_{1/2}$ (8.09 MeV) in the second one at 7.9 MeV.
The respective $N^{\star}_{\nu}$ and $N^{\star}_{\pi}$ indexes are lower than the corresponding ones in $^{28}$Ne.
The third peak at 10.5 MeV corresponds again to the low-energy peak of $^{26}$Ne
with again a similar microscopical structure: it is dominated (67.4\%) by the
$\nu$ 2s$_{1/2}$$\to$2p$_{3/2}$ (10.20 MeV) transition.
This peak is characterized by the same 2-qp configurations and the same transition densities as the $N=16$ isotones
and the neon isotopes $^{26}$Ne and $^{28}$Ne.
According to the value of
$N^{\star}_{\nu}$ it seems to be little more collective with respect to the corresponding excitation
in the other nuclei.
In the fourth peak at 12.8 MeV the main transition $\nu$ 1d$_{3/2}$$\to$1f$_{5/2}$ (12.08 MeV) (54.3 \%) is more mixed
to the $\nu$ 1d$_{5/2}$$\to$1f$_{7/2}$ (12.87 MeV) (18.6 \%) one. As in $^{28}$Ne, this fourth peak is again the more collective state with the typical isoscalar behavior of the transition densities.
The features of this fourth peak of $^{28}$Ne and $^{30}$Ne are more in touch with the standard representation of the PDR.

\subsection{Proton electric dipole resonance in $^{18}$Ne}

As can be observed in Fig. \ref{fig_dens}, the $^{18}$Ne exhibits a proton skin in spite of the presence of the Coulomb barrier.
It is interesting to investigate if in this case a low-energy proton electric dipole resonance appears, paying attention to its
degree of collectivity. Until now, only a few studies of dipole excitations in proton-rich nuclei have been done.
In a work based on relativistic QRPA on $N=20$ isotopes and $Z=18$ isotones \cite{Paar:2005fb}, it has been shown that proton pygmy dipole resonances may appear
when approaching the proton drip line. Here we focus on the $^{18}$Ne case. 
In this nucleus the proton Fermi level is 1d$_{5/2}$. In the conditions of our HFB ground-state calculations (9 HO major shells, 
absence of Coulomb exchange field) the occupation probability of this level is $v^2$=0.32 while the corresponding single particle energy 
is $\epsilon_{\pi\textrm{1d}_{5/2}} = -$2.14 MeV. The root-mean-square radius calculated for this orbital is 
$\sqrt{\langle r^2 \rangle_{\pi\textrm{1d}_{5/2}}}$ = 3.40 fm, to be compared to the total proton root-mean-square radius 
$\sqrt{\langle r^2 \rangle_{\pi}}$ = 2.84 fm.

\begin{figure}
\begin{center}
  \includegraphics[]{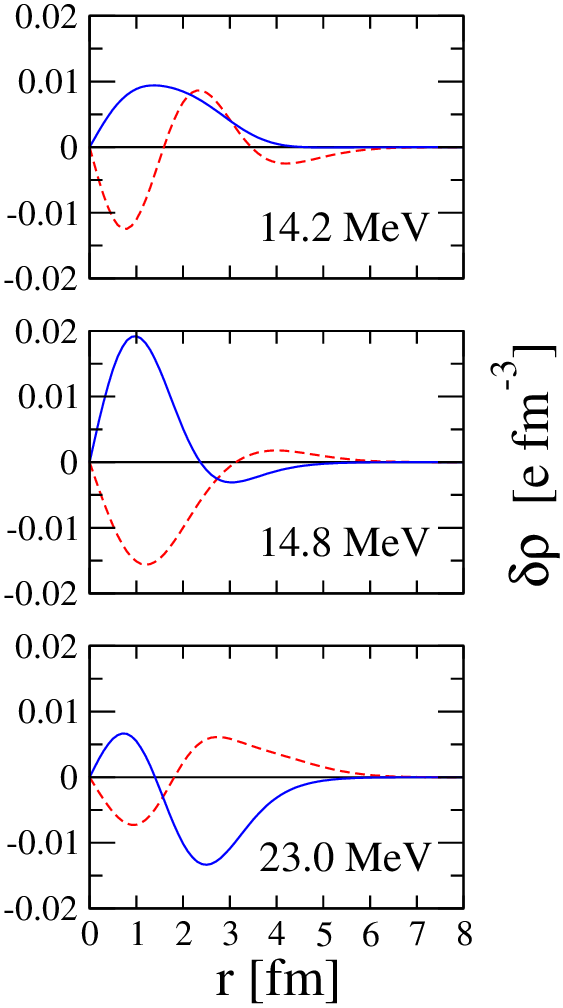}
\caption{(Color online) Neutron (full lines) and proton (dashed lines) transition densities for $^{18}$Ne.
The corresponding excitation energies are given in the panels. }
\label{fig_ne18_dens}
\end{center}
\vspace*{+0.8cm}
\end{figure}

Turning to the excited states, the first panel of Fig. \ref{fig_be1_z10} shows the $^{18}$Ne B(E1) 
distribution which is dominated by the GDR around 23 MeV. On the other hand, two distinct peaks appear below 15 MeV.
Their microscopical structures are given in Table \ref{table_18ne}. The first peak at 14.2 MeV is essentially due to proton excitations.
The dominant configurations correspond to transitions from the weakly bound proton state 1d$_{5/2}$. Contributions of 10\% order of magnitude arise from transitions from
1p$_{1/2}$, 2s$_{1/2}$ and 1p$_{3/2}$ proton states. The total neutron contribution is less than 8\%. The situation is different for the second low-energy peak,
located at 14.8 MeV, which is mainly due to neutron and proton 1p$_{1/2}$$\to$ 2s$_{1/2}$ transitions.
The behavior of the first peak, characterized by a superposition of many 2-qp proton configurations,
is specular with respect to one of the other neon isotopes and $N=16$ isotones: Only the roles of neutron and proton are obviously inverted.
This clearly appears also by looking at the transition densities
of Fig. \ref{fig_ne18_dens}: In this case, the proton and neutron are in opposite phase in the nuclear interior and in phase in the surface region, 
where the proton contribution
is larger than the neutron one. On the other hand, the 14.8 MeV excited state already displays a pronounced isovector behavior as the
GDR here appearing around 23 MeV. It is also very important to observe that the degree of collectivity of the first low energy state is
the same as the one of the GDR since the $N^{\star}$ values (shown in Table \ref{table_ne_isotop_Nstar_index}) are the same.
The $N^{\star}_{\pi}$ value is greater for the first low-energy peak revealing once again the dominant role of protons in this state. 
The correspondence between the results obtained for the low-lying excitation of this proton-rich nucleus and the neutron-rich nuclei 
here analyzed suggests that systematic studies of proton-rich nuclei in this mass region should be performed.

\section{Summary and conclusions}
We have studied the dipole excitations of neon isotopes and $N=16$ isotones focusing in particular on the low-excitation-energy region.
We performed a fully consistent axially-symmetric-deformed HFB+QRPA calculation using the D1S Gogny nucleon-nucleon effective force.
For the $N=16$ isotones we considered, we obtained  excitation modes between 9 and 12 MeV. All these modes presented a small but finite collective behavior:
also other transitions with respect to the dominant $\nu$ 2s$_{1/2}$$\to$2p$_{3/2}$ contribute to generate the low-lying resonance.
On the other hand, the spatial structure of the less-bound neutron state, that is, the $\nu$ 2s$_{1/2}$, is the main responsible of the neutron transition density
and this is reflected in the nontrivial macroscopic isoscalar or isovector behavior of the low energy resonance.
It holds also for the neon isotopes where the behavior of the transition densities is essentially due to spatial structure of the main hole state
contributing to the particle-hole transitions. Furthermore, also for the neon isotopes, we found that the strength of the low-energy state increases with the neutron-to-proton ratio.
By increasing the number of neutrons the dipole strength distribution becomes more fragmented. In $^{28}$Ne and $^{30}$Ne the occupation of the $\nu$ 1d$_{3/2}$ shell leads to the appearance of
another state just above the one corresponding to the excitation belonging  to all the $N=16$ nuclei analyzed. This state is more collective and
it is characterized by an isoscalar behavior of the transition densities, revealing features that are more in touch with the standard representation of the PDR characterizing heavier nuclei.

Finally, for the proton-rich $^{18}$Ne, we obtained a low-energy resonance at 14.2 MeV whose transition density behavior is quite similar to the one of the low-lying excitation of $N=16$ isotones
once the role of neutrons and protons is inverted. The degree of collectivity of this state is the same as the one of the main GDR peak of this nucleus.

\newpage

\begin{table}[t]
    \begin{center}
        \begin{tabular}{c|c|c|c|c}
            \hline
            $ $                                    & $^{24}$O                   & $^{26}$Ne                    &  $^{28}$Mg                        & $^{30}$Si                   \\
            \hline
            First peak                             &  $\omega_n$=9.1 MeV                   &   $\omega_n$=10.7 MeV                   &   $\omega_n$=11.6 MeV                       &   $\omega_n$=12.2 MeV                  \\
            $\nu$ 2s$_{1/2}$$\to$2p$_{3/2}$        & 73.5 \% (8.92 MeV)         &   67.6 \% (10.52 MeV)        &    39.7 \% (11.68 MeV)            &   43.8  \% (12.61 MeV)      \\
            $\nu$ 1d$_{3/2}$$\to$2p$_{3/2}$        &  0.0 \% (9.26 MeV)         &    2.8 \% (10.82 MeV)        &    32.7 \% (11.93 MeV)            &    7.0Ž  \% (12.79 MeV)      \\
            $\nu$ 2s$_{1/2}$$\to$2p$_{1/2}$        & 10.0 \% (10.33 MeV)        &    9.5 \% (12.44 MeV)        &     5.8 \% (13.98 MeV)            &    6.6Ž  \% (15.23 MeV)      \\
            $\nu$ 1d$_{5/2}$$\to$1f$_{7/2}$        &  8.7 \% (12.87 MeV)        &    8.7 \% (13.68 MeV)        &     8.2 \% (14.18 MeV)            &   11.1Ž  \% (14.55 MeV)      \\
            $\nu$ 1d$_{3/2}$$\to$2p$_{1/2}$        &  0.0 \% (10.67 MeV)        &    0.1 \% (12.74 MeV)        &     2.7 \% (14.22 MeV)            &    1.1Ž  \% (15.41 MeV)      \\
            $\nu$ 1p$_{1/2}$$\to$1d$_{3/2}$        &  1.2 \% (17.70 MeV)        &    1.1 \% (18.50 MeV)        &     0.5 \% (19.10 MeV)            &    0.3  \% (19.60 MeV)      \\
            $\nu$ total contribution               & 94.6 \%                    &   91.9 \%                    &    90.6 \%                        &   70.9  \%                  \\
            $\pi$ 1p$_{1/2}$$\to$2s$_{1/2}$        &  2.0 \% (15.19 MeV)        &    4.1 \% (15.97 MeV)        &     5.6 \% (16.22 MeV)            &   18.5  \% (15.92 MeV)      \\
            $\pi$ p$_{3/2}$ 1d$_{5/2}$             &  2.7 \% (15.89 MeV)        & 2.0 \% (17.60 MeV)           &  0.8 \% (18.97 MeV)               &    0.9  \% (17.15 MeV)\\
                                                   &(1p$_{3/2}$$\to$ 1d$_{5/2}$)& (1p$_{3/2}$$\to$ 1d$_{5/2}$) & (1p$_{3/2}$$\to$ 1d$_{5/2}$)      & (1d$_{5/2}$$\to$2p$_{3/2}$) \\
            $\pi$ 1d$_{5/2}$$\to$1f$_{7/2}$        &  0.0 \% (24.80 MeV)        &    0.9 \% (17.48 MeV)        &     1.9 \% (16.02 MeV)            &    7.7  \% (14.33 MeV)      \\
            $\pi$ total contribution               &  5.4 \%                    &   8.1  \%                    &     9.4 \%                        &   29.1  \%                  \\
            \hline
            Main peak (GDR)                        &   $\omega_n$=20.5 MeV                   &   $\omega_n$=21.9 MeV                   &    $\omega_n$=22.6 MeV                      &    $\omega_n$=23.1 MeV                 \\
            $\nu$ total contribution               &  80.5 \%                     &  64.5 \%                     &   78.5 \%                        &   54.4  \%                  \\
            $\pi$ total contribution               &  19.5 \%                    &  35.5  \%                    &   21.5 \%                        &   45.6  \%                  \\
            \hline
   Peaks with & B(E1) $\geq$ 0.5 e$^2$ fm$^2$& and  15 MeV & $\leq$ $\omega_n$ $\leq$& 30 MeV \\
            $\nu$ total contribution               &  81.6 \%                     &  78.0 \%                     &   82.9 \%                        &   52.7  \%                  \\   
            $\pi$ total contribution               &  18.4 \%                     &  22.0 \%                     &   17.1 \%                        &   47.3  \%                  \\
            \hline
            $N/A$                                  &  66.7 \%                   &    61.5 \%                     &   57.1 \%                        &   53.3  \%   \\
            \hline
        \end{tabular}
    \caption{
    \label{table1}
        Main particle-hole (and particle-particle) configurations contributing to the first low-energy dipole excitations of $N=16$ isotones.
     Total neutron and proton contributions are also given for the low-energy peak, for the main one corresponding to the GDR, 
and from the sum of all the states with 15 MeV $\leq \omega_n \leq$ 30 MeV and B(E1) $\geq$ 0.5 e$^2$ fm$^2$.}
    \end{center}
\end{table}

\begin{table}[t]
    \begin{center}
        \begin{tabular}{c c c c c c}
            \hline
            $\omega_n$ (MeV) & $N^{\star}$ & $N^{\star}_{\nu}$ & $N^{\star}_{\pi}$  &  \%($\nu$) & \% ($\pi$)                 \\
            \hline
           \textbf{$^{24}$O} & & & &  \\
           9.1      & 44& 34 & 10 &   94.6 \% & 5.4 \% \\
           20.5     & 74& 62 & 12 &   80.5 \% & 19.5 \% \\
           \textbf{$^{26}$Ne} & & & &  \\
           10.7      & 48& 30 & 18 & 91.9 \% & 8.1 \% \\
           21.9      & 82& 56 & 26 & 64.5 \% & 35.5 \% \\
           \textbf{$^{28}$Mg} & & & &  \\
           11.6      & 46& 24 & 22 & 90.6 \% & 9.4 \% \\
           22.6      & 114& 82 & 32 & 78.5 \% & 21.5 \% \\
           \textbf{$^{30}$Si} & & & &  \\
           12.2      & 54& 32 & 22 & 70.9 \% & 29.1 \% \\
           23.1      & 88&54 & 34 & 54.4 \% & 45.6 \% \\

            \hline
        \end{tabular}
    \caption{
    \label{table_N16_isoton_Nstar_index}
Values of the collectivity indexes for the low-lying dipole excitation and GDR in the $N=16$ isotones.}
    \end{center}
\end{table}

\begin{table}[t]
    \begin{center}
        \begin{tabular}{c c c c c c}
            \hline
            $\omega_n$ (MeV) & $N^{\star}$ & $N^{\star}_{\nu}$ & $N^{\star}_{\pi}$  & \%($\nu$) & \%($\pi$)             \\
            \hline
           \textbf{$^{18}$Ne} & & & &  \\
           14.2      & 72& 12 & 60 &   7.8 \% & 92.2 \% \\
           14.8      & 46& 10 & 36 &  44.3 \% & 55.7 \% \\
           23.0      & 72& 16 & 56 &  21.9 \% & 78.1 \% \\
           \textbf{$^{24}$Ne} & & & &  \\
           13.6      & 52& 44 & 8  &  97.7 \% & 2.3 \% \\
           16.4      & 48& 32 & 16 &  55.1 \% & 44.9 \% \\
           23.9      & 84& 60 & 24 &  65.0 \% & 35.0 \% \\
           \textbf{$^{26}$Ne} & & & &  \\
           10.7      & 48& 30 & 18 & 91.9 \% & 8.1 \% \\
           21.9      & 82& 56 & 26 & 65.5 \% & 35.5 \% \\
           \textbf{$^{28}$Ne} & & & &  \\
           8.1       & 36& 24 & 12 & 98.7 \% & 1.3 \% \\
           9.5       & 40& 26 & 14 & 98.4 \% & 1.6 \% \\
           11.1      & 46& 30 & 16 & 95.2 \% & 4.8 \% \\
           13.7      & 86& 72 & 14 & 97.1 \% & 2.9 \% \\
           16.1      & 78& 32 & 46 & 5.0  \% & 95.0 \% \\
           18.3      & 102& 76 & 26 & 91.5 \% & 8.5  \% \\
           20.9      & 124& 92 & 32 & 70.2 \% & 29.8 \% \\
           \textbf{$^{30}$Ne} & & & &  \\
           6.6       & 28& 18 & 10 & 98.9 \% & 1.1 \% \\
           7.9       & 34& 20 & 14 & 98.9 \% & 1.1 \% \\
           10.5      & 56& 40 & 16 & 96.9 \% & 3.1 \% \\
           12.8      & 80& 64 & 16 & 97.0 \% & 3.0 \% \\
           16.0      & 68& 26 & 42 & 3.2  \% & 96.8 \% \\
           18.7      & 142 & 98 & 44 & 68.3 \% & 31.7 \% \\
            \hline
        \end{tabular}
    \caption{
    \label{table_ne_isotop_Nstar_index}
    Values of the collectivity indexes for the main excitations in the neon isotopes.}
    \end{center}
\end{table}

\begin{table}[t]
    \begin{center}
        \begin{tabular}{c|c|c|c}
            \hline
            $ $                                    &   First low-energy peak &  Second low-energy peak &    Main peak (GDR)       \\
                                                   &  $\omega_n$=14.2 MeV & $\omega_n$=14.8 MeV &  $\omega_n$=23.0 MeV \\
             \hline
            $\pi$ 1d$_{5/2}$$\to$1f$_{7/2}$  (15.99 MeV)      & 29.1 \% &  1.6 \%  & \\
            $\pi$ 1d$_{5/2}$$\to$2p$_{3/2}$  (15.98 MeV)      & 23.2 \% &  6.3 \%  &\\
            $\pi$ 1p$_{1/2}$$\to$2s$_{1/2}$  (13.75 MeV)      & 11.7 \% &  35.2 \%  &\\
            $\pi$ 2s$_{1/2}$$\to$2p$_{3/2}$  (18.19 MeV)      &  9.3 \% &  1.2\%  &\\
            $\pi$ 1p$_{3/2}$$\to$1d$_{5/2}$  (16.95 MeV)      &  8.5 \% &  9.3 \% &\\
            $\pi$ 2s$_{1/2}$$\to$2p$_{3/2}$  (19.14 MeV)      &  2.7 \% &   0.5\% &\\
            $\pi$ total contribution                          & 92.2 \% & 55.7  \% & 78.1 \%   \\
& & &\\
            $\nu$ 1p$_{3/2}$$\to$1d$_{3/2}$  (15.78 MeV)      &  3.8 \% & 2.7 \%&   \\
            $\nu$ 1p$_{1/2}$$\to$ 2s$_{1/2}$ (14.18 MeV)      &  3.6 \% & 41.0 \%&   \\
            $\nu$ total contribution                          &  7.8 \% & 44.3 \% & 21.9 \%   \\
             \hline
        \end{tabular}
    \caption{
    \label{table_18ne}
             Main particle-hole (and particle-particle) configurations contributing to the first two low-energy peaks of $^{18}$Ne.
     Total neutron and proton contributions are also given both for the low-energy peaks and for the one corresponding to the GDR.}
    \end{center}
\end{table}


\begin{thebibliography}{0}
\bibitem {BOT37} W. Bothe, and W. Gentner, Z. Phys. {\bf 71}, 236 (1937).

\bibitem {ww} M.N. Harakeh, A. van der Woude,  {\it Giant Resonances:
Fundamental High-Frequency Modes of Nuclear Excitation} (Oxford University Press, Oxford, 2001).


\bibitem{Leistenschneider:2001zz}
  A.~Leistenschneider {\it et al.},
  Phys.\ Rev.\ Lett.\  {\bf 86}, 5442 (2001).


\bibitem{Ryezayeva:2002zz}
  N.~Ryezayeva {\it et al.},
  Phys.\ Rev.\ Lett.\  {\bf 89}, 272502 (2002).


\bibitem{Hartmann:2004zz}
  T.~Hartmann, M.~Babilon, S.~Kamerdzhiev, E.~Litvinova, D.~Savran, S.~Volz and A.~Zilges,
  Phys.\ Rev.\ Lett.\  {\bf 93}, 192501 (2004).


\bibitem{Adrich:2005zz}
  P.~Adrich {\it et al.},
  Phys.\ Rev.\ Lett.\  {\bf 95}, 132501 (2005).


\bibitem{Schwengner:2008rk}
  R.~Schwengner {\it et al.},
  Phys.\ Rev.\  C {\bf 78}, 064314 (2008).


\bibitem{Savran:2008zz}
  D.~Savran {\it et al.},
  Phys.\ Rev.\ Lett.\  {\bf 100}, 232501 (2008).

\bibitem{Gibelin:2008zz}
  J.~Gibelin {\it et al.},
  Phys.\ Rev.\ Lett.\  {\bf 101}, 212503 (2008).



\bibitem{Wieland:2009zz}
  O.~Wieland {\it et al.},
  Phys.\ Rev.\ Lett.\  {\bf 102}, 092502 (2009).



\bibitem{Tonchev:2010zz}
  A.~P.~Tonchev {\it et al.},
  Phys.\ Rev.\ Lett.\  {\bf 104}, 072501 (2010).



\bibitem{Paar:2007bk}
  N.~Paar, D.~Vretenar, E.~Khan and G.~Colo,
  Rept.\ Prog.\ Phys.\  {\bf 70}, 691 (2007).



\bibitem{Paar:2010ww}
  N.~Paar,
  J.\ Phys.\ G {\bf 37}, 064014 (2010).



\bibitem {gog1}J. Decharg\'e and D. Gogny, Phys. Rev. C {\bf 21}, 1568 (1980).
               J.F. Berger, M. Girod, and D. Gogny, Comp. Phys. Comm. {\bf 63}, 365 (1991).

\bibitem{Peru:2008gd}
  S.~P\'eru and H.~Goutte,
  Phys.\ Rev.\  C {\bf 77}, 044313 (2008).

\bibitem {u8} S.~P\'eru, G. Gosselin, M. Martini, M. Dupuis, S. Hilaire and J.-C. Devaux, 
Phys.\ Rev.\  C {\bf 83}, 014314 (2011).



\bibitem{Peru:2005di}
  S.~P\'eru, J.~F.~Berger and P.~F.~Bortignon,
  Eur.\ Phys.\ J.\  A {\bf 26}, 25 (2005).








\bibitem{Goriely:1998}
  S.~Goriely,
  Phys. Lett.  B {\bf 436}, 10 (1998).


\bibitem{Goriely:2002cx}
  S.~Goriely and E.~Khan,
  Nucl.\ Phys.\  A {\bf 706}, 217 (2002).

\bibitem{o24} V. Lapoux and H. Otsu, RIKEN NP0802-RIBF 57;
H. Baba, RIKEN NP0802-RIBF 56.

\bibitem{Matea} I. Matea, E611 - GANIL-PAC, March 2010, accepted proposal.

\bibitem{Carbone:2010az}
  A.~Carbone, G.~Colo, A.~Bracco, L.~G.~Cao, P.~F.~Bortignon, F.~Camera and O.~Wieland,
  Phys.\ Rev.\  C {\bf 81}, 041301 (2010).


\bibitem{Klimkiewicz:2007zz}
  A.~Klimkiewicz {\it et al.},
  Phys.\ Rev.\  C {\bf 76}, 051603 (2007).


\bibitem{Co':2009gi}
  G.~Co', V.~De Donno, C.~Maieron, M.~Anguiano and A.~M.~Lallena,
  Phys.\ Rev.\  C {\bf 80}, 014308 (2009).


\bibitem{Stephane}
http://www-phynu.cea.fr/HFB-Gogny\_eng.htm

\bibitem{Obertelli:2004qg}
  A.~Obertelli, S.~Peru, J.~P.~Delaroche, A.~Gillibert, M.~Girod and H.~Goutte,
  Phys.\ Rev.\  C {\bf 71}, 024304 (2005).


\bibitem{Cao:2005bt}
  L.~G.~Cao and Z.~Y.~Ma,
  Phys.\ Rev.\  C {\bf 71}, 034305 (2005).

\bibitem{Pena:2007}
D.~Pena Arteaga, P.~Ring,
Prog. Part. Nucl. Phys. {\bf 59}, 314 (2007).

\bibitem{Peru:2007}
  S.~P\'eru, H.~Goutte, J.~F.~Berger
  Nucl.\ Phys.\  A {\bf 788}, 44c (2007).

\bibitem{Yoshida:2008rw}
  K.~Yoshida and N.~Van Giai,
  Phys.\ Rev.\  C {\bf 78}, 014305 (2008).

\bibitem{Ebata:2010qr}
  S.~Ebata, T.~Nakatsukasa, T.~Inakura, K.~Yoshida, Y.~Hashimoto and K.~Yabana,
  Phys.\ Rev.\  C {\bf 82}, 034306 (2010).

\bibitem{Matsuo:2004pr}
  M.~Matsuo, K.~Mizuyama and Y.~Serizawa,
  Phys.\ Rev.\  C {\bf 71}, 064326 (2005).




\bibitem{Paar:2005fb}
  N.~Paar, D.~Vretenar and P.~Ring,
  Phys.\ Rev.\ Lett.\  {\bf 94}, 182501 (2005).








\end{thebibliography}
\end{document}